\documentclass[a4paper,fleqn]{cas-dc}
\usepackage[compress]{natbib}

\usepackage{pifont}
\usepackage{caption}
\usepackage{subfig}

\usepackage{times}
\usepackage{epsfig}
\usepackage{graphicx}
\usepackage{amsmath}
\usepackage{amssymb}
\usepackage{booktabs}
\usepackage{multirow}
\usepackage{pifont}
\usepackage{booktabs} 
\usepackage{multirow}
\usepackage{svg}

\def\tsc#1{\csdef{#1}{\textsc{\lowercase{#1}}\xspace}}
\tsc{WGM}
\tsc{QE}
\tsc{EP}
\tsc{PMS}
\tsc{BEC}
\tsc{DE}

\begin{document}
\let\WriteBookmarks\relax
\def\floatpagepagefraction{1}
\def\textpagefraction{.001}

\shorttitle{On the in vivo recognition of kidney stones using machine learning}

\shortauthors{F. Lopez-Tiro et~al.}

\title [mode = title]{On the in vivo recognition of kidney stones using machine learning}        

\author[addr1,addr2]{Francisco Lopez-Tiro}
\author[addr3]{Vincent Estrade}
\author[addr4,addr5]{Jacques Hubert}
\author[addr1]{Daniel Flores-Araiza}
\author[addr1]{Miguel Gonzalez-Mendoza} 
\author[addr1]{Gilberto Ochoa-Ruiz}
\author[addr2]{Christian Daul}

\cortext[cor]{Corresponding author \\ Email: gilberto.ochoa@tec.mx (G. Ochoa-Ruiz), christian.daul@univ-lorraine.fr (C. Daul*)}

\address[addr1]{Tecnologico de Monterrey, Escuela de Ingenieria y Ciencias, Av. Eugenio Garza Sada Sur 2501 Sur, Tecnológico, 64849 Monterrey, N.L, Mexico}

\address[addr2]{Centre de Recherche en Automatique de Nancy (UMR 7030, CNRS and Université de Lorraine), 2 avenue de la For\^et de Haye, F-54516 Vand{\oe}uvre-L\`es-Nancy, France}

\address[addr3]{CHRU Pellegrin, place Am\'elie Raba L\'eon, F-33000 Bordeaux, France}

\address[addr4]{IADI-UL-INSERM (U1254), 5 rue du Morvan, Vand{\oe}uvre-l\`es-Nancy, France}

\address[addr5]{CHRU Nancy, Service d’urologie de Brabois, rue du Morvan, F-54511 Vand{\oe}uvre-L\`es-Nancy, France}

\maketitle
\begin{abstract}
Determining the kidney stones type allows urologists to prescribe a treatment to avoid recurrence of renal lithiasis. An automated in-vivo image-based classification method would be an important step towards an immediate identification of the kidney stone type required as a first phase of the diagnosis. In the literature it was shown on ex-vivo data (i.e., in very controlled scene and image acquisition conditions) that an automated kidney stone classification is indeed feasible. This pilot study compares the kidney stone recognition performances of six shallow machine learning methods and three deep-learning architectures which were tested with in-vivo images of the four most frequent urinary calculi types acquired with an endoscope during standard ureteroscopies. This contribution details the database construction and the design of the tested kidney stones classifiers. Even if the best results were obtained by the Inception v3 architecture (weighted precision, recall and F1-score of 0.97, 0.98 and 0.97, respectively), it is also shown that choosing an appropriate colour space and texture features allows a shallow machine learning method to approach closely the performances of the most promising deep-learning methods (the XGBoost classifier led to weighted precision, recall and F1-score values of 0.96).
\end{abstract}



\section{Introduction}
%

\begin{table*}[!h]
\centering
\begin{tabular}{|l|l|l|l|}
\hline
\bf{Type} & \bf{Acronym} & \bf{Occurrences} & \bf{Crystal shape} \\ \hline
Type I = Whewellite  & WW & from 15 up to 35 \% & Dumbbell shape \\ \hline
Type II = Weddellite & WD & from 15 up to 35 \% & Bipyramidal \\ \hline
Type IIIb = Uric Acid dihydrate & UA & from 2 up to 13 \%& Flat rhomboidal \\ \hline

Type IVc = Struvite & STR & from 20 up to 30 \% & Rectangular Prism \\ \hline
Type IVd = Brushite & BRU & from 5 up to 20 \%& Elongate narrow \\ \hline
Type V = Cystine & CYS & from 1 up to 3 \%& Hexagonal plates \\ \hline
\end{tabular}
\caption{Simplified urinary calculi classification:  the frequency of appearance and a description of the crystal morphology are given for each stone type.}
\label{ks_class}
\end{table*}
Urinary lithiasis refers to the formation of crystalline accretions (kidney stones) from minerals dissolved in urine \cite{cloutier2015kidney}. Kidney stones form themselves in the kidneys and migrate through the urinary tract (ureters, bladder, etc.). While small kidney stones evacuate naturally and imperceptibly, larger accretions (beyond a few millimeters) often cause severe pain (e.g., due to an obstructed ureter) and must be removed during an ureteroscopy (endoscopy of the upper urinary tract). Numerous developed countries~\cite{kasidas2004renal, hall2009nephrolithiasis} exhibit a high urinary lithiasis incidence since about 10\% of their population is affected at least once by a kidney stone episode. The formation of kidney stones is favoured by various risk factors. Apart from reasons related to genetic inheritance, diet (eating too many fruits, vitamin C, vitamin B6, or animal proteins increases the risk of forming kidney stones), chronic diseases (e.g., diabetes) or an inappropriate lifestyle (e.g., a sedentary lifestyle that leads to a high body mass index) are some of the risk factors for urinary lithiasis. There is a direct relationship between these risk factors and the biochemical composition of the kidney stones~\cite{silva2010chemical, daudon2012stone}. In developed countries, the stone recurrence rate approaches a very high value of 40\%~\cite{scales2012prevalence, viljoen2019renal}. 

Therefore, identifying the kidney stone types is crucial to avoid relapses~\cite{kartha2013impact, friedlander2015diet} through personalized treatments (diet adaptation, surgery, etc.) is considered of utmost importance by many practitioners~\cite{estrade2013place}. For this purpose, several guidelines for visually recognizing some of the more common types of kidney stones (see Table ~\ref{ks_class}) have been proposed in recent years~\cite{daudon2004clinical, estrade2017pourquoi} to be applied in the clinical practice.

The international morpho-constitutional classification of urinary stones includes seven groups denoted by roman numerals going from I to VII. As seen in Table 1, each group is associated with a specific crystalline type. Groups I to V designate whewellite, weddellite, uric acid dihydrate, calcium and non-calcium phosphates (i.e., brushite) and cystine, respectively.  Group VI contains protein rich calculi which are very infrequent and group VII gathers all other kidney stone types. Each group is itself divided into several subgroups to differentiate morphologies and aetiologies for a given crystalline type (the subgroup names are designated by letters which complete the roman numbers). The most recent lithogenic events (i.e., cristal type) are located on the surface, whereas less recent events are observable on the section~\cite{corrales2021classification}. 

\subsection{Context and recent trends in ureteroscopy}

Kidney stones~\cite{basiri2012state} can be detected using various techniques, such as computed tomography~\cite{kawahara2016predicting},  the ultrasound modality or  X-ray imaging~\cite{hidas2010determination, mccarthy2016radiology}. Kidney stones can be destroyed using ``extracorporeal shock wave lithotripsy’’ or ultrasound ureterolithotripsy~\cite{singh2014kidney}. However, the most widespread kidney stone diagnosis and removal technique is ureteroscopy, which is typically associated with laser lithotripsy~\cite{alenezi2015flexible}.

\begin{figure*}[!ht]
\centering
\includegraphics[width=0.75 \linewidth]{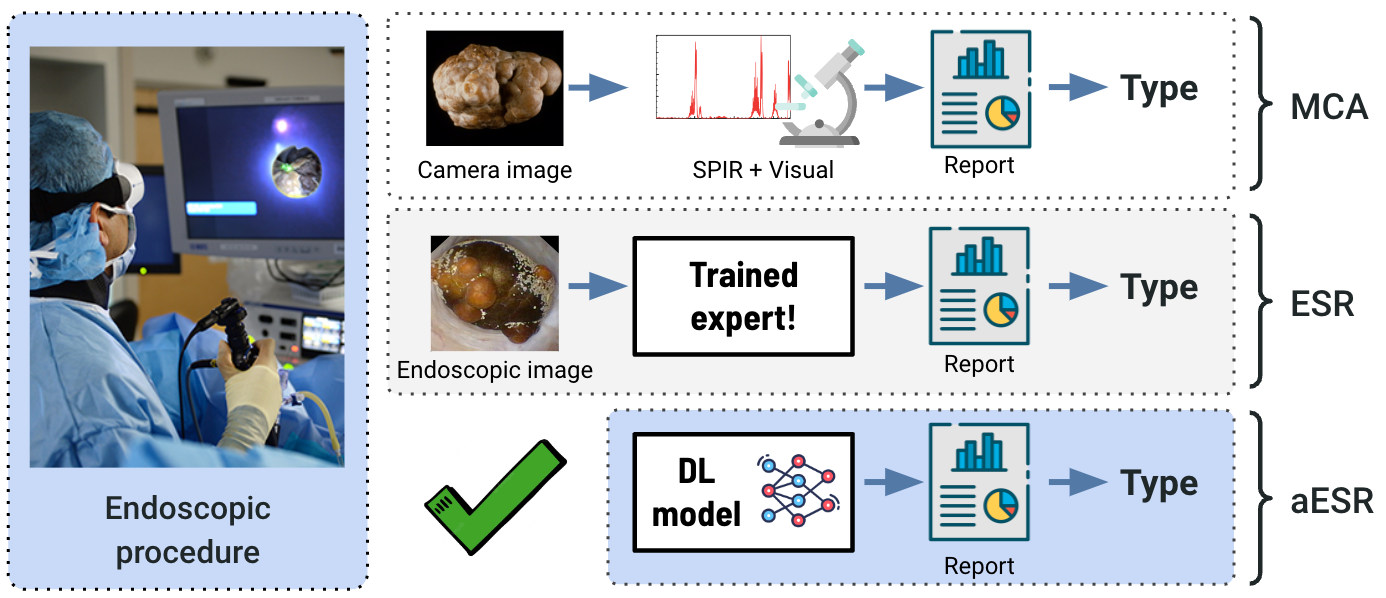}
\caption{Morpho-Constitutional Analysis (MCA, \cite{daudon2016comprehensive}) proposed by Michel Daudon is the standard guideline for the identification of kidney stones (top row). MCA performs a visual inspection and is complemented by a biochemical analysis on extracted kidney stones (post-surgery). Endoscopic Stone Recognition (ESR, \cite{estrade2013place}) is a technique proposed by Vincent Estrade to perform in-vivo kidney stone identification during surgery using only the information displayed on the screen (middle row). Automatic Endoscopic Stone Recognition (aESR) is a method based on computer vision techniques and machine learning to classify in-vivo endoscope images (bottom row).}
\label{fig:identification}
\end{figure*}

\begin{figure}[t]
\centering
\includegraphics[width=\linewidth]{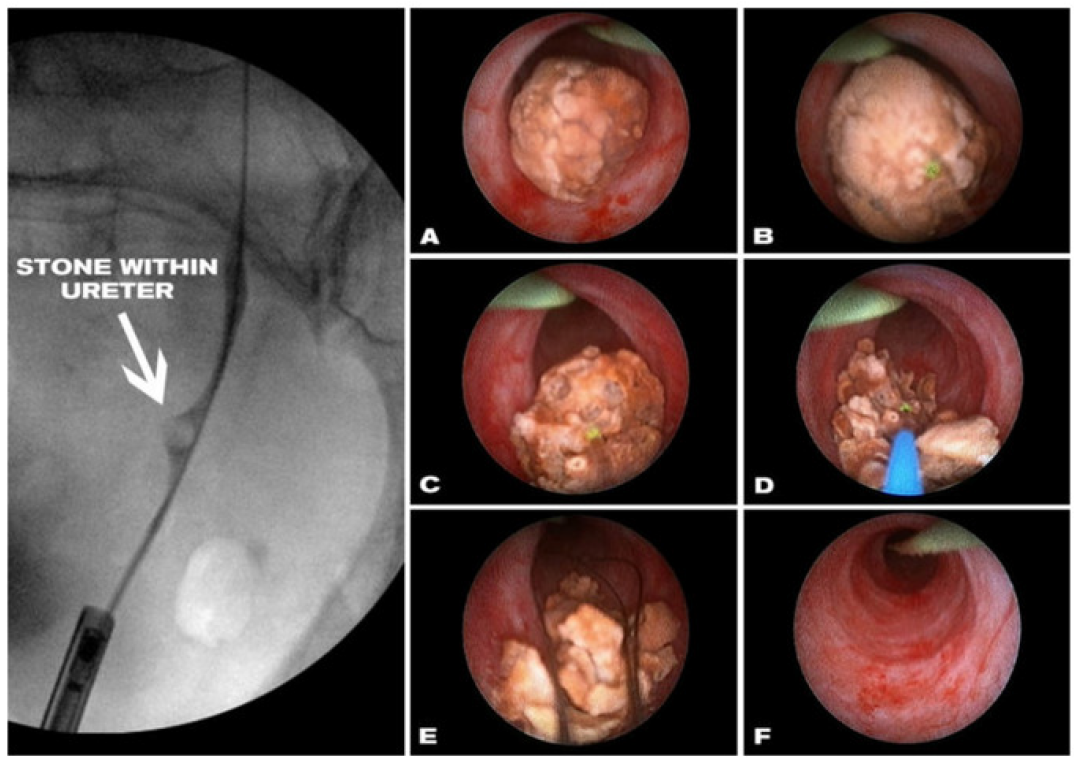}
\caption{Classical ureteral stone removal process. Left: a stent is introduced in the ureter to guide the introduction of the endoscope  (fluoroscopy image). A: Complete calculus visualized using an ureteroscope. B: Ureteral stone targeting using a laser (green dot). C and D: Ureteral stone fragmentation. E: Removal of stone fragments with a basket. F: Stone-free ureter.}
\label{fig:laser_process}
\end{figure}

Modern ureteroscopes are flexible endoscopes that illuminate the scene (the inner surfaces of hollow organs) with white light~\cite{castaneda2016evolution, de2017history}. The short focal length (approximately 6 mm) of the endoscope optics allows for the acquisition of high resolution images. The endoscope’s distal tip is close to the observed surface, while the short focal length ensures a rather large angle of view. In the ``chip on the tip’’ technology, the sensor (CCD) matrix is fixed on the distal tip and the electrical signals are transferred to the digital camera located on the proximal tip. A low-frequency (3-5 Hz) laser is inserted in the endoscope’s operating channel and is used to fragment and remove the kidney stones from the urinary tract (the fragmentation process is depicted on  Fig.~\ref{fig:laser_process}). The kidney stone fragments must remain large enough to allow for a morpho-constitutional analysis (MCA, top row of Figure \ref{fig:identification}) carried out with a microscope and for the subsequent infrared-spectrophotometry analysis~\cite{daudon2016comprehensive}. The visual examination under the microscope aims to define the kidney stone surface and section in terms of textures, appearance of the crystals, colours, and morphological particularities to fully assess the possible causes of lithogenesis. On the other hand, infrared-spectrophotometry enables to identify the molecular and crystalline composition of the different areas (layers) of the kidney stone. This full morpho-constitutional analysis allows to identify the type (or class) of the kidney stone and to prescribe appropriate treatments (surgery, diets, etc.) for minimizing recurrences.

\begin{table*}[h]
\centering
\arrayrulecolor{black}
\begin{tabular}{!{\color{black}\vrule}l!{\color{black}\vrule}l!{\color{black}\vrule}l!{\color{black}\vrule}l!{\color{black}\vrule}l!{\color{black}\vrule}l!{\color{black}\vrule}l!{\color{black}\vrule}l!{\color{black}\vrule}l!{\color{black}\vrule}l!{\color{black}\vrule}} 
\hline
\multirow{2}{*}{Reference/Feature} & \multicolumn{6}{l!{\color{black}\vrule}}{Kidney Stone Composition} & \multicolumn{2}{l!{\color{black}\vrule}}{Image Type} & \multirow{2}{*}{Acquisition}  \\ 
\cline{2-9}
                                   & AU         & WW         & WD          & STR         & CYS           & BRU           & Surface        & Section  &          \\ 
\hline
{}\cite{serrat2017mystone}              & \ding{51}  & \ding{51}   & \ding{51}    & \ding{51}   &  &                  & \ding{51}    &  \ding{51}        &      Ex vivo     \\ 
\cline{1-1}
{}\cite{torrell2018metric}             & \ding{51}  & \ding{51}   & \ding{51}    & \ding{51}   &     & \ding{51}   & \ding{51}      &           &    Ex vivo        \\ 
\cline{1-1} 
{} \cite{black2020deep}              & \ding{51}  & \ding{51}   &              & \ding{51}   & \ding{51}      & \ding{51}    & \ding{51}      & \ding{51}  &   Ex vivo          \\ 
\cline{1-1}
{} \cite{martinez2020towards}               &         \ding{51}  & \ding{51}   &  \ding{51}       &   &     &    &  \ding{51}    & \ding{51}  &   In vivo          \\ 
\cline{1-1}
{} \cite{estrade2021toward}               &         \ding{51}  & \ding{51}   &  \ding{51}       &   &     &    &  \ding{51}    & \ding{51}  &   In vivo          \\ 
\cline{1-1}
This contribution                               & \ding{51}  & \ding{51}   & \ding{51}     &            &               & \ding{51}    & \ding{51}       &         \ding{51}        &           In vivo           \\
\hline
\end{tabular}
\arrayrulecolor{black}
\caption{Overview of the kidney stone classes and acquisition conditions in the-state-of-the-art works, as well as for this contribution.  Simplified taxonomy: UA: Uric Acid (Anhydrous and Dihydrate), COM: Calcium Oxalate Monohydrate (whewellite, WW), COD: Calcium Oxalate Dihydrate (weddellite, WD), STR: Struvite, CYS: Cystine, BRU: Brushite.}
\label{pworks1}
\end{table*}

However, kidney stone identification using a morpho-constitutional analysis has two major drawbacks. First, in numerous hospitals, the results of this widespread analysis are often available only after one or two months, even if in critical situations a therapeutic decision should immediately be taken. Second, removing fragments of kidney stones is a tedious and time-consuming task that can last from 30 up to 60 minutes. In this context, the increase in the imaging quality of endoscopes is leading more and more urologists seek to visually identify the morphology (or crystalline type) of kidney stones only with the help of images displayed on a screen. To do so, kidney stones are first fragmented in large parts, and both their external surfaces and their sections are visually analysed. After their identification, the fragmented  kidney stones are either collected or vaporized using a laser lithotripsy technique called “dusting”. Fragmentation and dusting can be performed with the same laser, the difference being in the settings of the instrument. In the dusting mode, low energy (0.2-0.5 J) and high frequency (10-20 Hz) are used to vaporize the kidney stones~\cite{de2019metabolic}.

A visual analysis performed by an urologist can be an appropriate first step for identifying  the crystal type of a kidney stone. However, such an analysis requires a great deal of experience due to the high inter-class similarities and intra-class variations of the stones. This visual analysis, which is designated by the term ``endoscopic stone recognition'' (ESR,~\cite{estrade2022towards}, middle row of Figure \ref{fig:identification}),  can currently only be achieved by a limited number of specialists, whereas the management of urolithiasis diseases is part of the daily life of every urologist in the world. Moreover, even for experienced specialists, the classification remains often operator-dependent~\cite{siener2016quality, sampogna2021identificacion}. Therefore, the implementation of automated and reproducible classification methods in this context would make it possible to take full advantage of the dusting technique.

\begin{table*}[h]
\begin{tabular}{|l|c|c|c|c|c|c|c|c|}
\hline
\multirow{2}{*}{Reference} & \multicolumn{6}{c|}{Precision Per Class} & \multirow{2}{*}{\begin{tabular}[c]{@{}c@{}}Weighted \\ Precision\end{tabular}} & \multirow{2}{*}{ML Method} \\ \cline{2-7}
 & AU & WW & WD & STR & CYS & BRU &  &  \\ \hline
{}\cite{serrat2017mystone}{} & 0.65 & 0.55 & 0.69 & 0.50 & N/A & N/A & 0.63 & Random Forest \\ \hline
{}\cite{torrell2018metric}{} & 0.76 & 0.67 & 0.80 & 071 & N/A & 0.72 & 0.74 & Siamese CNN \\ \hline
{}\cite{black2020deep}{} & 0.94 & 0.95 & N/A & 0.71 & 0.75 & 0.75 & 0.85 & CNN – ResNet101 \\ \hline
\cite{martinez2020towards}& 0.91 & 0.94 & 0.92 & N/A & N/A & N/A & 0.92 & Random Forest \\ \hline
\cite{estrade2021toward} & 0.99 & 0.90 & 0.93 & N/A & N/A & N/A & 0.94 & ResNet152v2 \\ \hline
\end{tabular}
\caption{Comparison of the  precision obtained by the three main contributions for the most common kidney stone classes. The precision is given for each individual class and classifier. The taxonomy per stone class same as in Table \ref{pworks1}. The average precision (weighted by the image number of each class) are also given for each kidney stone type. In the last table line, the precision values are given for surface images. For section images, the authors in \cite{estrade2021toward}, reported a precision of 0.95, 94 and 0,94 for AU, WW and WD, respectively. The corresponding weighted precision equals 0.94.}
\label{pworks2}
\end{table*}

\subsection{Previous attempts of kidney stone classification}
\label{previous-attempts}

 Different approaches have been proposed to deal with the classification of kidney stones. These works exploit various image modalities such as hyper-spectral imaging~\cite{blanco2012hyperspectral}, non-contrast computer tomography~\cite{kawahara2016predicting, grosjean2013pitfalls} or multimodal data~\cite{kazemi2018novel}. However, these data modalities are not used in a standard clinical situation since only endoscopic images are visualized in urology departments. Table~\ref{pworks1} gathers the previous works which have addressed the problem of automatic endoscopic stone recognition (aESR, bottom row of Figure \ref{fig:identification}) using only images. This table provides also an overview on the used data (urinary calculus classes and acquisition conditions).

The first work~\cite{serrat2017mystone} dedicated to the classification of kidney stone images used ex-vivo data. In this contribution, collected kidney stones were placed in a closed enclosure that included light sources and cameras whose positions (distances and viewing angles) are optimized to capture large surface and section parts of fragmented kidney stones of known type. Images were acquired using a RGB CMOS 5 megapixel camera by switching between white and infrared light sources. Texture and color information was encoded in feature vectors to describe the kidney stones seen in high-resolution images. These feature vectors consist of RGB colour histograms and texture histograms of local and rotation invariant binary patterns. A Random Forest (RF) classifier was used to recognize the kidney stone type. Various ablation tests were performed to determine the best feature vector (e.g., the most informative features) and the best RF model configuration. Although the average accuracy of this method was rather moderate (see Table~\ref{pworks2}), the results showed that texture and colour information can potentially be discriminant enough to automate the classification of various types of kidney stones.

In a continuation of this precursor work, the classification results were improved in~\cite{torrell2018metric} by exploiting a deep learning technique. However, the choice of the neural network was dictated by the limited size of the dataset. Despite this drawback, the encouraging results (an average accuracy of 74\%, see Table~\ref{pworks2}) showed the potential of convolutional neural network (CNN) based methods to tackle this problem.

Another deep learning approach was also used in a recent study~\cite{black2020deep} to perform an ex-vivo classification. High resolution images were acquired for sixty-three kidney stones of various bio-chemical compositions (see the five stone classes given in Table~\ref{pworks2} for~\cite{black2020deep}) provided by a laboratory performing morpho-constitutional analyses. At least one image was acquired for both the surface and the section of each kidney stone fragment, so that two or more images were available for all samples. As in other previous works, patches (including only kidney stone parts) were extracted from the images and fed to the machine learning algorithm.

In this work, a deep CNN, namely ResNet-101 which was pre-trained with ImageNet to prevent overfitting was used for feature extraction and classification of each patch of the dataset. The obtained model was assessed using the leave-one-out cross validation method, with the primary monitored outcome being the model recall to account for the reduced size of their dataset. While the average precision over five classes is the best for this contribution, the classification metrics remain rather low for some classes, see the struvite (STR), cystine (CYS), and the brushite (BRU) classes in Table~\ref{pworks2}.

A more recent effort  aiming to improve the classification of kidney stones images using traditional features (color histograms and LBP features for texture descriptors) and shallow methods (a Random Forest classifier) was presented Martinez et al.~\cite{martinez2020towards}. Their results showed that a well chosen color space can have a significant impact on the classification accuracy of 3 classes, attaining a 92\% weighted average accuracy for kidney stones using features from surface and sections patches, as shown in the fourth row of Table~\ref{pworks2}.

A last work~\cite{estrade2022towards} has investigated the applicability of deep CNNs to predict the morphology and composition of both pure and mixed stones. The authors made use of a dataset consisting of 347 images of the surface and section of kidney stones acquired with a Olympus URF-V flexible ureteroscope. However, it has to be noticed that, in contrast to most works in the recent literature, the authors also investigated the performance of machine learning models trained with a database of images including kidney stones with  pure crystalline composition and urinary calculi with several layers with different crystalline composition. In order to train and test their models, the authors used image patches with a size of $256 \times 256$ pixels and trained two multi-class classification models based on ResNet-152-V2 (one for surface images and another for section images). For kidney stones with a pure crystalline composition, the authors made use of the same dataset as the one used in~\cite{martinez2020towards} (see the fifth row of Table~\ref{pworks1}).  The authors in~\cite{estrade2022towards} expanded the training dataset using data augmentation to make their models more general. 

A cross-validation was repeated ten times with randomly chosen image combinations for the training and testing steps. The full process was also repeated with different random initialization seeds for the deep CNN algorithm. Average standard test metrics were reported for each step (i.e., precision, area under the ROC curve (AUROC), specificity, sensitivity, see Table~\ref{concordance}). Compared to other works in the literature~\cite{martinez2020towards}, the authors in~\cite{estrade2022towards} did not investigate the effect of mixing sections and surface features during the network training, which has shown to improve the overall kidney stones recognition task.

\begin{table*}[!t]
\center
\begin{tabular}{|l|l|l|l|l|l|l|}
\hline
Component & Stone type & \begin{tabular}[c]{@{}l@{}}\% Correct Matches\end{tabular} & \begin{tabular}[c]{@{}l@{}}Number of    \\ occurrences\end{tabular} & AUC & \begin{tabular}[c]{@{}l@{}}Precision\\  (PPV)\end{tabular} & \begin{tabular}[c]{@{}l@{}}Recall\\(Sensitivity) \end{tabular} \\ \hline
Whewellite & Ia or Ib & 85\% & 205 & 0.87 & 0.88 & 0.85 \\ \hline
Weddellite & IIa or IIb & 85\% & 178 & 0.87 & 0.86 & 0.85 \\ \hline
Uric Acid & IIIa or IIIb & 91\% & 64 & 0.95 & 0.94 & 0.91 \\ \hline
Carbapatite & IVa & 81\% & 176 & 0.86 & 0.88 & 0.81 \\ \hline
Struvite & IVc & 50\% & 10 & 0.74 & 0.45 & 0.50 \\ \hline
Brushite & IVd & 65\% & 23 & 0.82 & 0.83 & 0.65 \\ \hline
Cystine & Va & 100\% & 7 & N/A & N/A & N/A \\ \hline
\end{tabular}
\caption{Matched results between endoscopic and microscopic studies. ``N/A'' stands for not applicable  and refers to data which are insufficient for a statistical use. This simplified class representation was adapted from \cite{estrade2021toward}}
\label{concordance}
\end{table*}

The best sensitivity was obtained for the type IIIb (uric acid) using surface images (98\% of the IIIb kidney stone images were correctly predicted). The most frequently encountered morphology was the Ia type (pure whewellite composition). It was correctly predicted in 91\% and 94\% of the cases using surface and section images, respectively.

All these preliminary and promising results described above explain why the medical community in urology is convinced of the interest of kidney stone recognition methods based on artificial intelligence~\cite{fitri2020automated} and of the importance of incorporating computer aided diagnosis tools in their workflow~\cite{jahrreiss2020artificial}. This work is an extension of a preliminary study~\cite{martinez2020towards} which still improved the classification results using a RF classifier. However, the high precision of this last work was obtained for 3 classes only (see Table~\ref{pworks2}).

\subsection{Objectives and structure of the paper}

Except in~\cite{martinez2020towards, estrade2022towards}, the kidney stone images used in the existing methods described in Section \textit{``Previous attempts of kidney stone classification''} were acquired in ex-vivo under controlled acquisition conditions (well defined acquisition viewpoints, large and contrasted kidney stone surfaces, high resolution images, and diffuse illumination without reflections on the surfaces). These contributions gave an indication about the feasibility of in vivo kidney stone classification in optimal conditions and could be used to automate and speed-up the morpho-constitutional analysis, even if the described algorithms were probably not conceived with this aim.

In vivo images acquired with flexible uteroscopes are far of being captured in optimal conditions. On the one hand, it is difficult to control the endoscope’s position and thus optimal acquisition angles and distances cannot be warranted. Large kidney stone surfaces and sections are difficult to be systematically captured. Moreover, in the vast majority of urology centres, ureteroscopes are equipped by cameras with HD sensor matrices (1024 x 720 pixels) whose resolution is clearly smaller than 5 mega-pixels used in previous works. Compared to the images used in existing the literature, the acquired kidney stone fragments are much smaller and with lower resolution. On the other hand, images can suffer from numerous artefacts. For instance, motion blur due to high and non-constant endoscope displacement speeds and defocusing/refocusing due to changing distal tip/kidney stone distances affect globally the quality of numerous images. Moreover, specular reflections are also often visible due to the crystalline nature of the kidney stones and floating objects may occlude some regions of interest. Even if numerous efforts have been made in endoscopy to detect and segment artefacts in various applications (as in urology, gastroscopy or colonoscopy, see~\cite{trinh2018mosaicing, ali2020objective}), images that are automatically selected in endoscopic videos are not often of optimal quality.

The aim of this contribution is to demonstrate the feasibility of the classification of kidney stones acquired in vivo during ureteroscopic procedures. The first results of a previous contribution~\cite{martinez2020towards} indicate that shallow (or classical) machine learning algorithms can improve the results of the literature, even when in vivo images are used. One goal of this contribution is to delve deeper into the feature selection and classifier tuning to improve the recognition results using shallow machine learning methods. This paper also investigates the advantages of using deep-learning methods for the classification of in vivo data and compare their performances to those of shallow machine learning approaches. 
The paper is organized as follows. Section \textit{``Materials and Methods’’} describes the construction of three steps, namely the collection of in vivo images, the optimal patch extraction from the images, and the tested class balancing methods. The first part of Section \textit{``Design of a set of kidney stone recognition methods''} describes the main aspects of shallow machine learning methods (feature extraction, optimal feature selection, and the classifier tuning) for an automated kidney stone recognition. The second part of Section \textit{``Design of a set of kidney stone recognition methods''} deals with the presentation of several deep-learning approaches for kidney stone classification. Section \textit{``Results and Discussion''} details the results obtained by the different solutions described in Section \textit{``Design of a set of kidney stone recognition methods''} and compares them with the results obtained in the literature. A  conclusion and perspectives are given in Section \textit{``Conclusions''}. 
\section{Database construction \label{dataset_desc}}
\begin{table*}[!t]
\centering
\begin{tabular}{|c|c|c|c|c|}
\hline
\multicolumn{2}{|c|}{\textbf{Image Type}} & \multicolumn{2}{c|}{\textbf{Acquired Images}} &\textbf{Number} \\ \cline{1-4}
\textbf{View} & \textbf{Class} & \textbf{Number} & \textbf{Occurrence (\%)} & \begin{tabular}[c]{@{}c@{}}\textbf{of patches}\end{tabular} \\ \hline
\multirow{5}{*}{\textbf{Surface}} & WW (Type Ia) & 30 & 31.9 & 870 \\ \cline{2-5} 
 & WD (Type IIb) & 32 & 34.1 & 920 \\ \cline{2-5} 
 & UA (Type IIIb) & 18 & 19.1 & 470 \\ \cline{2-5} 
 & BRU (Type IVd) & 14 & 14.9 & 420 \\ \cline{2-5} 
 & \textbf{Total} & \textbf{94} & \textbf{100.0} & \textbf{2680} \\ \hline
\multirow{5}{*}{\textbf{Section}} & WW (Type Ia) & 27 & 31.0 & 820 \\ \cline{2-5} 
 & WD (Type IIb) & 28 & 32.2 & 780 \\ \cline{2-5} 
 & UA (Type IIIb) & 18 & 20.7 & 460 \\ \cline{2-5} 
 & BRU (Type IVd) & 14 & 16.1 & 410 \\ \cline{2-5} 
 & \textbf{Total} & \textbf{87} & \textbf{100.0} & \textbf{2470} \\ \hline
\end{tabular}
\caption{Number of acquired images and of their (almost) non overlapping square patches. The whewellite (Type Ia), wedellite (Type IIb), uric acid (Type IIIb) and brushite (Type IVb) classes include 57 (30 surface and 27 section) images, 60 (32 + 28) images, 36 (18 + 18) images and 28 (14 + 14) images, respectively.}
\label{patches_qty}
\end{table*}

The images used in this contribution (see Table~\ref{patches_qty}) were acquired by an urologist (Dr. Vincent Estrade) who is among the few experts in France able to visually recognize kidney stone types using only in vivo images displayed on a screen during an ureteroscopy. Additionally to this expertise, the annotation of the images used in this work was statistically confirmed in~\cite{estrade2021toward} by a concordance study exploiting the morpho-constitutional analysis of extracted kidney stone fragments (the morpho-constitutional analysis confirmed the visual classification made by the urologist), using microscopy and a  Fourier transform infrared spectroscopy (FTIR) analysis, which were exploited as follows.

The ESR  started with a visual observation of the kidney stone surfaces. Then, the kidney stones were split in two fragments using a laser.  The Holmium-Yag laser parameters were set as follows: the laser pulse frequency, energy and power were adjusted at 5 Hz, 1.2–1.4 J and 6–7 W, respectively.  The pulse sequence length was short and the fibre diameter was either 230 or 270 $\mu m$. A second visual observation of the fragment section was then performed. An additional fragmentation session was carried out when needed to allow for the analysis all types of pure and mixed stones.

The goal of the concordance study in~\cite{estrade2022towards} was to assess the efficiency of the ESR process. To do so, the fragmented kidney stones were analysed by a biologist  (a MD with 40 years of experience) which performed  a morphological analysis (visual inspection under a microscope), and a FTIR analysis done with a spectrometer. In a standard morpho-constitutional analysis, the surface,  the section and the nucleus of each kidney stone are inspected. In the concordance study in~\cite{estrade2022towards}, the visual recognition of kidney stone types was considered as ``confirmed'' when a non-significant difference between the endoscopic and microscopic images (p-value $>$ 0.05) was obtained. The area under the ROC curve (AUROC), the sensitivity, the specificity, the positive predictive value (PPV) and the negative predictive value (NPV) were calculated for a dataset with a confirmed concordance. The authors of~\cite{estrade2022towards} regrouped the following pairs of kidney stone morphologies that have similar aetiologies: Ia and Ib, IIa and IIb, IIIa and IIIb, Va and Vb, and VIa and VIb. These six morphological types cover 95\% of the most common pure stones that urologists encounter in their daily practice.  

Indeed, despite the variety of compositions and morphologies observed in urinary calculi, about 90\% are composed of a limited number of crystalline species and morphological characteristics
that are easily recognised through an endoscopic examination~\cite{estrade2013place, estrade2017pourquoi}. The study in~\cite{estrade2022towards} showed a high concordance between endoscopic and microscopic typing of the kidney stones. More precisely, the concordance of the results was observed in 86.1\% of the type I kidney stones, in 85\% of the type II kidney stones, in 91\% of the type III kidney stones and in 79\% of type IV kidney stones made of calcium phosphate. Table~\ref{concordance} gives more details about this concordance.

A urologist (Vincent Estrade, 20 years of experience) intra-operatively and prospectively collected endoscopic digital images
and videos of stones used for this study between January 2018 and November 2020 in a single centre (CHU Pellegrin, Bordeaux). A flexible digital ureterorenoscope (Olympus URF-V CCD sensor) was employed. The study adhered to all local regulations and data protection agency recommendations (National Commission on Data Privacy requirements).
Patients were informed that their data would be used anonymously.

\begin{figure}[hbt]
\centering
\includegraphics[width=\linewidth]{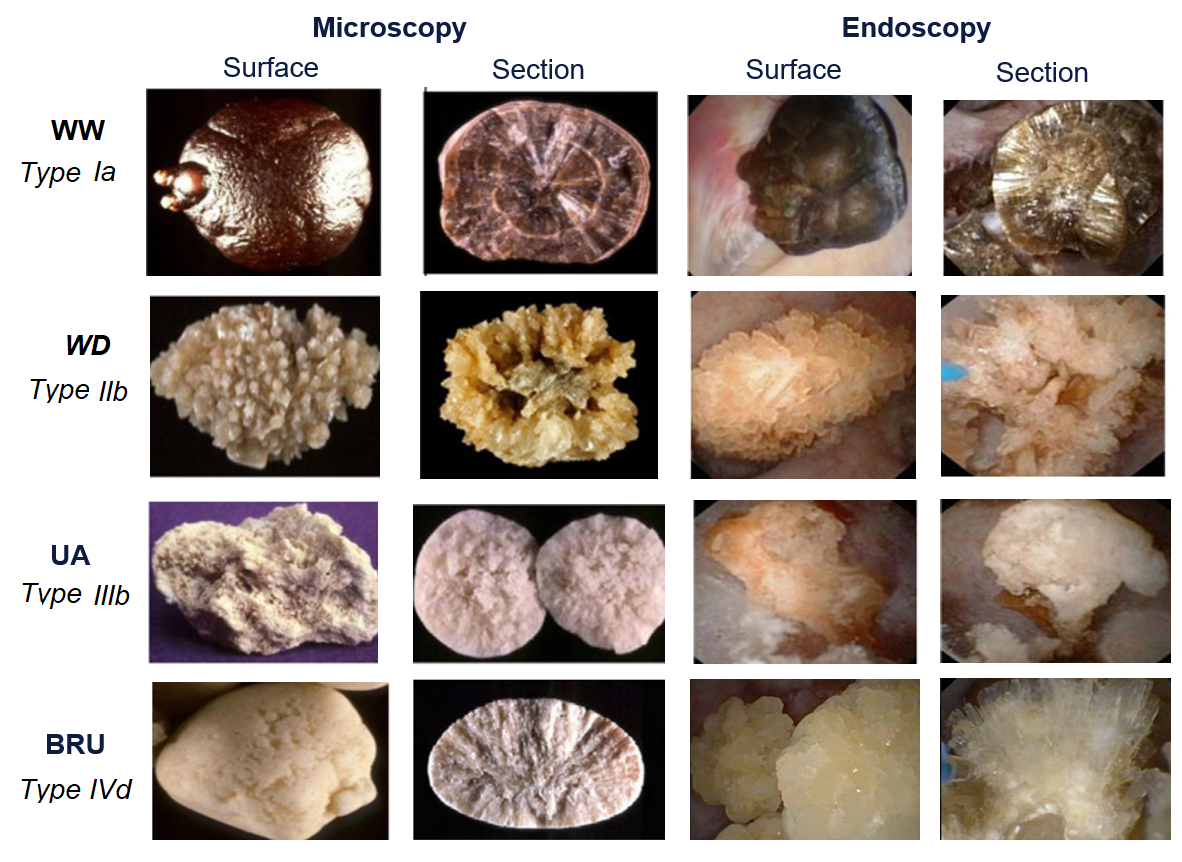}
\caption{Ex-vivo surface (first column) and section (second column) images acquired with a microscope, and in vivo surface (third column) and section (fourth column) images captured with an endoscope. It has to be noticed that the kidney stones are not the same for the two image modalities. 
When moving from the first to the last line one have successively following classes: WW (Type Ia), WD (Type IIb), uric acid (Type IIIb) and brushite (Type IVd).}
\label{fig:stone_types}
\end{figure}

\subsection{Image dataset \label{image_set_desc}} 

The dataset includes 181 kidney stone images which were acquired with four ureteroscope models, namely two from the Olympus company (the URF-V and URF-V2 endoscopes), and two other models from the Richard Wolf company (two different BOA models).The use of different endoscopic devices increases the variability of the image quality due to the acquisition conditions (changing illumination, uncontrolled viewpoints, etc.). Kidney stone surface and section images are shown for each of the four classes in the last two columns of Fig.~\ref{fig:stone_types}. The relative image count of the images of the four classes (see Table~\ref{patches_qty}) is in accordance with the typical kidney stone type occurrences observed in clinical situation (see Table~\ref{ks_class}). The classes consist of 57 (31\%),  60 (33\%), 36 (19.9\%) and 28 (13.3\%) images for the WW, WD, acid uric and brushite kidney stone types, respectively.

Experts who visually analyse kidney stone types do not observe globally a complete image, but rather interpret the image content by successively exploiting texture and color information of several image regions. The experts interpret in this way both the microscope images (see the two columns on the left of Fig.~\ref{fig:stone_types}) and the in vivo images (see the two columns on the right of Fig.~\ref{fig:stone_types}) observed during a morpho-constitutional analysis and an ureteroscopic procedure, respectively. For this reason, the classification of urinary calculi is not performed on whole images in~\cite{serrat2017mystone, martinez2020towards}, but using square patches localized on kidney stone surfaces or sections delineated by edges (surrounding tissues should not be visible in the patches). In this contribution, stone fragment edges are automatically segmented in the images using a learning-based active contour method~\cite{hatamizadeh2020end}.

\subsection{Patch extraction} \label{patch_section}

However, in some images, patches may contain parts of instruments used to fragment or extract kidney stones. These patches are identifiable since the instruments are easy to segment in the blue channel of the images (contrary to the instruments, epithelial tissues and kidney stones are characterized by color values for which the red and green channels carry stronger signals than the blue channel).          

Thus, three precautions have to be taken during the patch extraction. First, the extracted patches have a maximal border overlap of twenty pixels to limit redundant information. Second, patches including a very high number of ``non-kidney stone'' pixels are not included in the dataset (an experimentally set threshold value of 10\% was used to discard inappropriate patches located close to the fragment periphery or including instruments). Third, the patches should have an optimal size to capture local texture and color information (such a size adjustment was not presented in previous works~\cite{serrat2017mystone, martinez2020towards}).    

\begin{figure}[t] 
    \centering
    
    \subfloat[]{
    \includegraphics[width=0.49\linewidth]{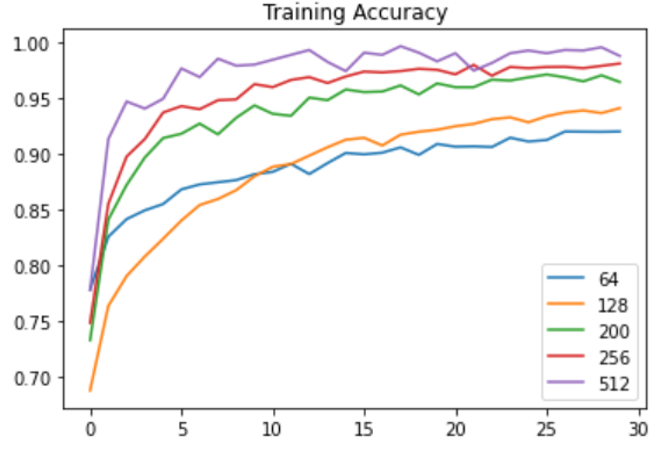}}
    \subfloat[]{
    \includegraphics[width=0.49\linewidth]{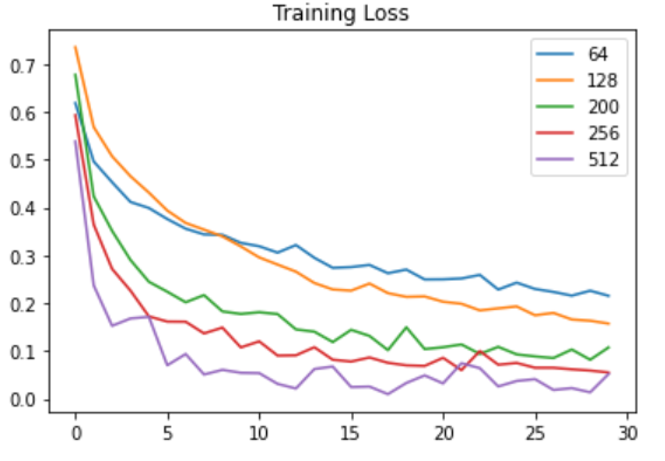}}

        \caption{Illustration of the classification efficiency dependence with the patch size. (a) Accuracy obtained for each patch size with a random forest tree and all 40 feature components (see Section \ref{feature_sel} for the handcrafted feature description). (b) Loss curves for the different patch sizes.}
        \label{fig:patch_sizes}
    \end{figure}

The side length of the square patches was a hyper-parameter which was adjusted during the training of the machine learning models presented in Section \textit{``Design of a set of kidney stone recognition methods''}. The best size value was obtained after several ablation studies using five patch areas ($64\times64$, $128\times128$, $200\times200$, $256\times256$ and $512\times512$ pixels, respectively) and by monitoring the training precision and loss curves for each patch size, as shown in Fig.~\ref{fig:patch_sizes} for a RF classifier. As noticeable in this figure, the accuracy and the loss values respectively increase and decrease significantly when the patch area becomes larger. 

Increasing the patch size beyond $512\times512$ pixels does not improve the performances of the classifiers. With a patch size of $512\times 512$  pixels, the accuracy of the shallow machine learning models is slightly better than with patches of $256\times 256$ pixels. However, with a patch area of $512\times 512$ pixels, the training of the deep learning models lead to over-fitting which is observable in the validation set.  Consequently, the  same $256\times256$  patch size was used for all the results reported in Sections \textit{``Design of a set of kidney stone recognition methods''} and \textit{``Results and discussion''}. With this procedure, 2680 and 2470 patches were respectively obtained for the surfaces and the sections of the complete kidney stone fragment database. The last column of Table~\ref{patches_qty} gives the number of patches for each class.
\subsection{Class balancing strategies. \label{unbalanced}}
The sizes and the viewpoints of the fragmented kidney stone surfaces and sections are very variable (i.e., the number of useful patches extracted from the images depend on the number of pixels which effectively correspond to urinary calculi). Moreover, the number of images per class is statistically dependant on the urinary calculi type (see Table~\ref{ks_class}). These two facts explain why the number of patches per class is imbalanced, as noticeable in Table~\ref{patches_qty}. 

Two strategies were tested to balance the classes. 
\begin{itemize}
    \item{\textbf{Over-sampling approach}.} For this strategy, the class with the highest patch number is taken as reference for the over-sampling. The WW class (Type Ia or Ib) is this reference class, for which 870 and 820 patches were extracted from the images of the surface and section fragments, respectively. The patch number of the brushite, WD and AU classes is increased by randomly  extracting additional patches (still with $256\times256$ pixels) which do not correspond  to the cells of the initial (regular) grid of patches. After this over-sampling, the four classes consist of 870 surface and 820 section patches.
    \item{\textbf{Under-sampling approach}.} This strategy follows a similar principle as the over-sampling since the number of patches of three classes (WW, WD and UA) is randomly reduced so that all urinary calculi classes  include the same patch amount as the class with the smallest number of patches (the BRU class consists of 420 surface and 410 section patches).    
\end{itemize}
These class balancing experiments were carried out with the Scikit Learn Imbalanced library~\cite{lemaitre2017imbalanced}.  Classification tests have shown that the ``up-sampling approach’’ led to slightly better classification accuracy, which means that the presence of redundant information (partially superimposed patches) is completely compensated by the increase in the number of patches. 
\section{Design of a set of kidney stone recognition methods. \label{proposed-method}}
\begin{figure*}[!t]
\centering
\includegraphics[width=0.7\linewidth]{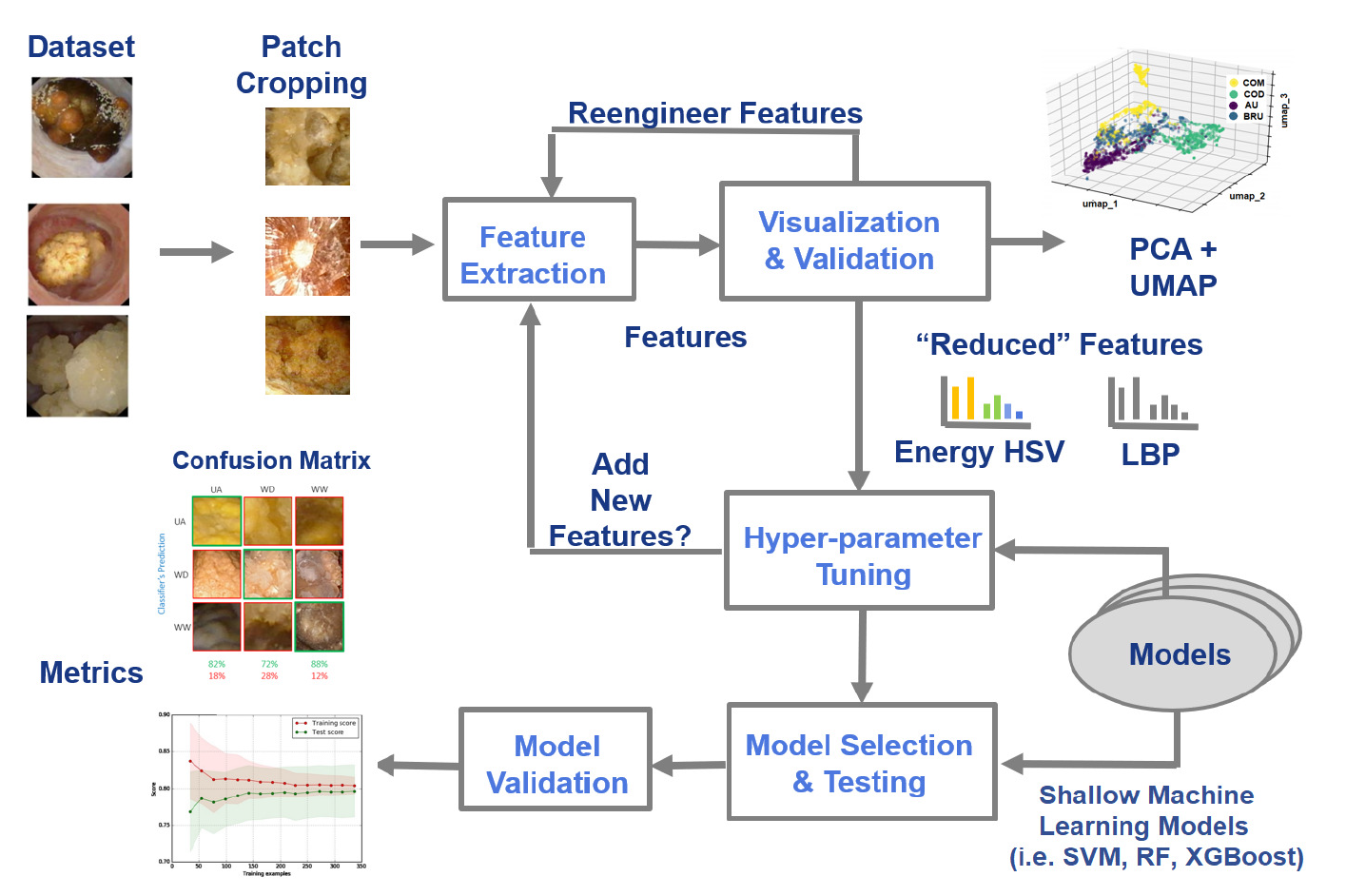}
\caption{Illustration of the training of the six shallow machine learning models:  feature extraction and selection, hyper-parameter tuning,  and model validation. }
\label{fig:diagram_shallow}
\end{figure*}

This section describes the training and validation of shallow and deep learning-based models. The training of the shallow machine learning models follows the traditional pipeline of Fig.~\ref{fig:diagram_shallow}, which is thoroughly discussed in Section \textit{``Shallow machine learning methods''}. For the deep learning-based methods a different approach is followed by training three well-known models using an end-to-end approach and transfer learning to compensate for the relatively small size of the available dataset, as depicted in  Fig.~\ref{fig:diagram_deep}. The different choices in terms of deep-learning model design, training of the networks and their validation are detailed in Section \textit{``Deep-learning based classification methods''}.

\subsection{Shallow machine learning methods. \label{shallow_methods}}
An overview of the training and validation of the shallow models is shown on Fig.~\ref{fig:diagram_shallow}, which can be summarized as follows. In Section \textit{``Database construction’’} it was detailed how the relatively small number of patches was increased for training both classical and deep learning models. Section \textit{``Handcrafted feature extraction and selection''} starts with an argued feature extraction process which encodes handcrafted  color and texture information in vectors.
The classification accuracy using different feature combinations was monitored to highlight the discriminating capacity of the features associated with a reference classifier. 
In~\cite{martinez2020towards} it was shown that random forest (RF) trees are appropriate to test the classification  efficiency according to different feature combinations. These tests were performed using a k-fold cross validation approach and the ability of the features to form separate clusters in the feature space is visualized using UMAP~\cite{mcinnes2018umap}. 
After the determination of the most discriminant feature vectors, various well known shallow models were trained using an iterative tuning of their hyper-parameters based on a k-fold cross validation, as thoroughly discussed in Section \ref{tuning_shallow}.
\subsubsection{Handcrafted feature extraction and selection.\label{feature_sel}}  
Previous works~\cite{serrat2017mystone, torrell2018metric, martinez2020towards} have shown that color and texture features are appropriate to describe the kidney stone surfaces and sections. This section goes more deeply in the justification of various choices made for the feature extraction (size of the local window for the texture encoding, appropriateness of the chosen color space, and color feature type) for training shallow machine learning models. 
%
%
\paragraph{\textbf{Color Features}\label{colFeatures}} 
Color spaces can be sorted in three families according to their general advantages~\cite{daul2000building}. The family of tri-stimulus spaces based on a set of primaries gathers all $RGB$ color spaces (including the $XYZ$ space). These colour spaces are widespread for technical reasons since cameras and screens acquire and display $RGB$ values, respectively. Numerous classification algorithms use $RGB$ color values which exhibit a major drawback: two visually different colors may be separated by a small distance in the color space coordinate system, and vice versa. For this reason, a second family of colors spaces (e.g. the $Lab$ and $Luv$ color spaces) were designed so that small or large numerical distances correspond to small and large color perception differences, respectively. These color spaces, in which the $L$ component stands for the color intensity and the two other components are chromaticity values, are optimized to measure color differences. However, they do not reproduce the color perception of the human brain which separates the hue and intensity information (the brain perceives more subtle hue differences than intensity changes). The third family of color spaces (e.g., $HSI$, $HSV$, etc.) is based on three components: hue (the tint information), the saturation (the ``amount of grey’’ in the color) and the brightness (the colour intensity).These color spaces allow for a closer simulation of the colour perception by the human brain since a change in only the light intensity does not affect the hue values. In the field of endoscopy, where the intensity of illumination can significantly change from one image to another, it is important to have an information (here the hue) that is independent of the image acquisition conditions. Thus, the $HSV$ space was used to extract the color information from the ureteroscopic images. The brightness, saturation and hue values are defined by~\eqref{HSVequ1},~\eqref{HSVequ2} and~\eqref{HSVequ3}, respectively,  where $MIN = min(R,G,B)$ and $H$ is in degrees.
\begin{equation} \label{HSVequ1}
V = max(R,G,B)
\end{equation}
\begin{equation} \label{HSVequ2}
S = \left\{\begin{matrix}
1-MIN/V & \textrm{if}\: \:  V \neq 0  \\ 
0 & \textrm{otherwise}
\end{matrix}\right.
\end{equation}
\begin{equation} \label{HSVequ3}
H = \left\{\begin{matrix}
 60(G-B)/(V-MIN) \:\: \textrm{if} \:\:  V=R\\ 
 120+60(B-R)/(V-MIN) \:\: \textrm{if} \:  V=G \\
 240+60(R-G)/(V-MIN)) \:\: \textrm{if} \:  V=B \\
 \textrm{Undefined}  \quad \textrm{if} \quad  R=G=B
\end{matrix}\right.
\end{equation}
After the $RGB$ to $HSV$ color conversion, the energies $e_{I_c}(x, y)$ were pixel-wise determined in each channel $I_c$ ($c$ = $H$, $S$ or $V$) using~\eqref{energyEq} in order to capture local color changes: 

\begin{equation}  
\left\{
\begin{matrix} \label{energyEq}
    e_{I_c}(x, y)  & = & \hspace*{-3.5mm}\sqrt{
     g_x\left(I_c(x,y)\right)^2 + g_y\left((I_c(x,y)\right)^2
      } \\
    g_x(I_c(x,y)) \hspace*{-3mm} & = & \hspace*{-3mm} I_c(x+1, y) - I_c(x-1, y) \\
    g_y(I_c(x,y)) \hspace*{-3mm} & = & \hspace*{-3mm} I_c(x, y+1) - I_c(x, y-1)
\end{matrix}
\right. 
\end{equation}
where $g_x(I_c(x, y))$ and $g_x(I_c(x, y))$ are gradient components along the $x$ and $y$ image axes. Representing the occurrences of these local energies using histograms lead to a global color description at the patch level. The energy values computed for each $I_c$ channel were used to build a ten-bin histogram (30 bins in all for the three channels).

\paragraph{\textbf{{Texture features}}} 
Haralick features determined with co-occurrence matrices are popular texture descriptors which are appropriate for large regions or entire images. The kidney stone fragment textures are strongly changing according to their location in the images, so that rotation invariant local binary pattern (LBP) values were preferred to capture texture information. These patterns were stored in histograms representing statistical information about local textures. Similarly to the optimal patch size search, classification tests were performed to find the best side size of the LBP windows. These window sizes (i.e., $5 \times 5$,  $7 \times 7$ and  $9 \times 9$) were hyper-parameters in the classification scheme depicted in Fig. 4. Using the RF classifier, the most discriminant texture features were obtained for a window area of $5 \times 5$ pixels. The LBP values are computed using grey level patches (the grey level values are given by the intensity channel of the $HSV$ space presented in section \ref{colFeatures}) and are used to determine 10 bin histograms. 

\paragraph{\textbf{{Texture and color feature vectors}}}  
The complete feature vector (extracted either from surface or section patches) consists of a 40-bin histogram encoding hue energies ($eH$), saturation energies ($eS$), intensity energies ($eV$) and texture ($LBP$) information. The RF classifier was used to test the discrimination capacity of the features taken individually ($eH$, $eS$, $eV$ or LBP taken separately, see Table \ref{rf_results}), partially combined in different ways ($eH$ + $eS$+ $eV$ or $LBP$ + $eH$) or all jointly used ($LBP$ + $eH$ +$eS$ + $eV$).
%
The 80 components of the feature vectors extracted from the images of the mixed dataset result from the concatenation of the 40 histogram bins of a surface patch and of that of a section patch of the same kidney stone fragment.   

\begin{table*}[t]
\centering
\begin{tabular}{|l|c|c|c|c|c|c|c|c|c|}
\hline
\multirow{2}{*}{\textbf{Descriptor}} & \multicolumn{3}{c|}{\textbf{Surface Images}} & \multicolumn{3}{c|}{\textbf{Section Images}} & \multicolumn{3}{c|}{\textbf{Mixed}} \\ \cline{2-10} 
 & Precision & Recall & F1 & Precision & Recall & F1 & Precision & Recall & F1 \\ \hline
\textit{eH} & 0.79 & 0.76 & 077 & 0.65 & 0.64 & 0.64 & 0.76 & 0.76 & 0.76 \\ \hline
\textit{eS} & 0.59 & 0.57 & 0.57 & 0.61 & 0.61 & 0.60 & 0.64 & 0.64 & 0.64 \\ \hline
\textit{eV} & 0.57 & 0.66 & 0.61 & 0.60 & 0.59 & 0.59 & 0.63 & 0.63 & 0.62 \\ \hline
\textit{eHSV} & 0.81 & 0.77 & 0.70 & 0.71 & 0.72 & 0.71 & 0.83 & 0.83 & 0.82 \\ \hline
\textit{LBP} & 0.85 & 0.85 & 0.85 & 0.73 & 0.72 & 0.71 & 0.88 & 0.87 & 0.88 \\ \hline
\textit{LBP + eH} & 0.87 & 0.86 & 0.86 & 0.75 & 0.75 & 0.75 & 0.89 & 0.88 & 0.88 \\ \hline
\textit{LBP + eHSV} & 0.87 & 0.86 & 0.86 & 0.83 & 0.82 & 0.83 & 0.91 & 0.91 & 0.91 \\ \hline
\end{tabular}
\vspace*{-1mm}
\caption{Random Forest performance using different feature descriptors for section, surface and mixed patches.} 
\label{rf_results}
\end{table*}
It is noticeable in Table \ref{rf_results} that, in contrast with the results reported in the previous works  (see also Table \ref{pworks2}), the LBP features associated with a RF tree yield relatively high accuracy and recall for both surface and section images. It is also noticeable that the hue channel taken separately provides significant discriminant capabilities. When using all texture and color features  together one reach a precision of 0.87 and 0.83 for surface and section images, respectively. 
\begin{figure*}
    \centering
    
    \subfloat[]{
    \includegraphics[width=0.35\linewidth]{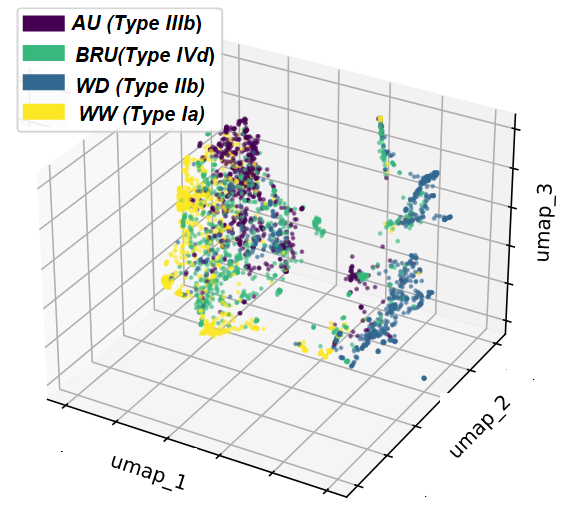}}
    \subfloat[]{
    \includegraphics[width=0.35\linewidth]{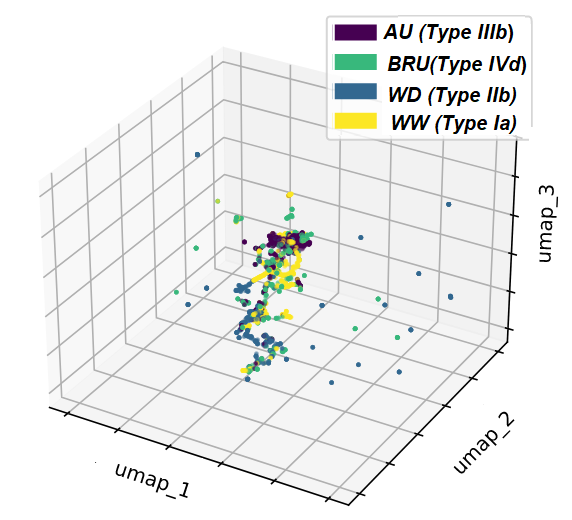}}
    \vspace*{-2mm}
    \subfloat[]{
    \includegraphics[width=0.35\linewidth]{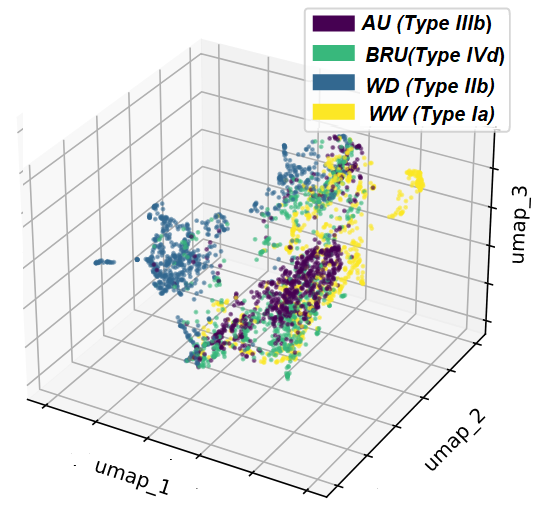}}
    \subfloat[]{
    \includegraphics[width=0.35\linewidth]{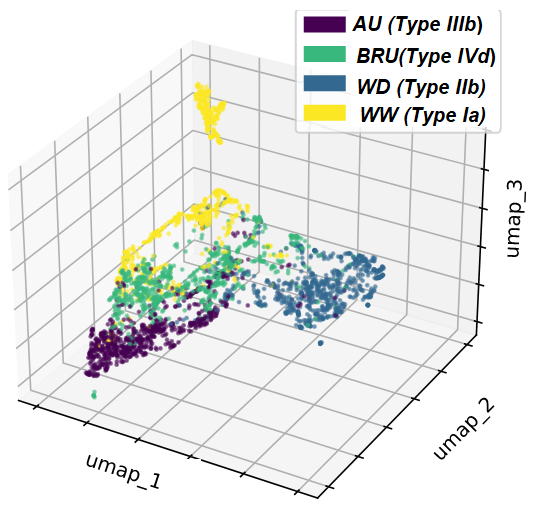}}
        \caption{UMAP visualisation for the representation of the class separability according to different feature combinations and the separate or simultaneous use of surface and section patches. This UMAP representation is achieved using only the three most discriminant dimensions (umap1 to umap3) obtained after a dimensionality reduction of the $HSV$-$LBP$ feature space. (a) Vector representation including all handcrafted $eH$,  $eS$, $eV$ and $LBP$ features extracted from surface images. (b) Same representation as in (a), but for section images. (c) UMAP feature space representation obtained when using only the colour information (the $2\times 30$ component vector of $eH$,  $eS$, $eV$ values) extracted both from section and surface images. (d) Same as in (c) but with all features ($eH$, $eS$, $eV$ and $LBP$) extracted from both patch types and encoded in 80-component vectors.}
        \label{fig:umap}
    \end{figure*}

UMAP visualizations, as given in Fig.~\ref{fig:umap}, are very helpful for understating the results obtained with the RF classifier exploiting the different feature combinations given in Table~\ref{rf_results}. Several interesting conclusions can be drawn from the plots in Fig.~\ref{fig:umap}.  Fig.~\ref{fig:umap}(a) shows that, for the surface patches,  the vector including all $HSV$ and $LBP$ feature histograms provides a high separability of the classes. This visual representation of the class separability is in accordance with the high precision of 0.87 obtained by the RF model. %
As shown in Fig.~\ref{fig:umap}(b), using only section images leads to a poorer class separability. This result is in accordance with the  visual classification by human experts reported in the concordance study given in~\cite{estrade2022towards} and is also confirmed in Table~\ref{rf_results} by the lower precision of 0.83.  

The hue component carries significant information leading to discernible clusters and, combining hue and texture descriptors (for both sections and surfaces) helps the classifier to attain a precision of 0.76 (see Fig.~\ref{fig:umap}(c)).
Finally, the clusters become more discernible when using the 80-component feature vectors including both surface and section information. The visualisation in  Fig.~\ref{fig:umap}(d) is again in accordance with the precision of 0.91 obtained by the RF classifier (see Table~\ref{rf_results}). 

It is noticeable in Table~\ref{rf_results} that, in contrast with the results reported in the previous works  (see also Table~\ref{pworks2}), the LBP features associated with a RF tree yield relatively high accuracy and recall for both surface and section images. It is also noticeable that the hue channel taken separately provides significant discriminant capabilities. When using all texture and color features  together one reach a precision of 0.87 and 0.83 for surface and section images, respectively. 

\begin{figure} [t!] 
    \centering
    
    \subfloat[]{
    \includegraphics[height=0.48\linewidth]{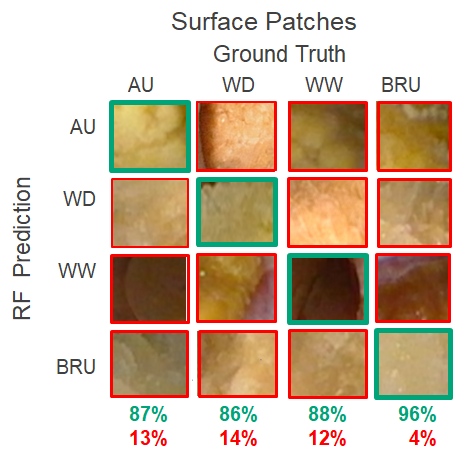}}
    \subfloat[]{
    \includegraphics[height=0.48\linewidth]{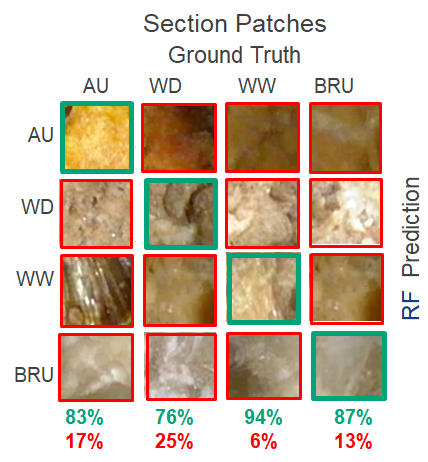}}

        \caption{Confusion matrix for the four classes of the dataset. The results were obtained with  a RF classifier (the percentages in green indicate the rate of correct recognition).  Classification  errors (percentages in red) are due to the intra-class variabilies and inter-class similarities}
\label{fig:conf_matrix}
    \end{figure}

Furthermore, it is also noticeable that, in comparison to the preliminary work in~\cite{martinez2020towards} which also uses a RF classifier, the increase on the number of classes (i.e., the introduction of brushite) was not done in detriment of the overall precision (0.92 in~\cite{martinez2020towards} and 0.91 in Table~\ref{rf_results}) obtained when using the 80-component feature vectors. Fig.~\ref{fig:conf_matrix} shows a confusion matrix (including representative patches for each class) that partially explain why shallow features are somewhat ineffective: the data exhibit high intra-class variations, such that the aspect of the images of two classes can be close in terms of color and texture.

\subsubsection{Tuning of the classifiers exploiting handcrafted features. \label{tuning_shallow}}
As depicted in Fig.~\ref{fig:diagram_shallow}, after having identified the best combination of handcrafted features, a set of six state-of-the-art machine learning models was trained (see Table~\ref{ml_results}).  The hyper-parameter tuning consisted of  a combination of a grid and random search  using the Scikit-Learn software~\cite{lemaitre2017imbalanced}. This search of optimal hyper-parameters was performed for three datasets, namely i) by using only the surface patches from which vectors gathering all 40 color and LBP components were extracted,  ii) by exploiting  the same vectors extracted only from the section patches and iii) by exploiting the 80 component vectors obtained with a surface and section patch. The last dataset is the most representative of the clinical practice since the aspect of both the surfaces and the sections are taken into account by when human operators visually identify kidney stones. The two other datasets are useful to assess the contribution of the surface and section information taken separately. A stratified k-fold cross validation approach was used in order to maximize the number of data in the testing phase and to mitigate biases. The precision, recall and F1-scores were measured for each of the three dataset configurations.

 \begin{figure*}[!t]
\centering
\includegraphics[width=0.7\linewidth]{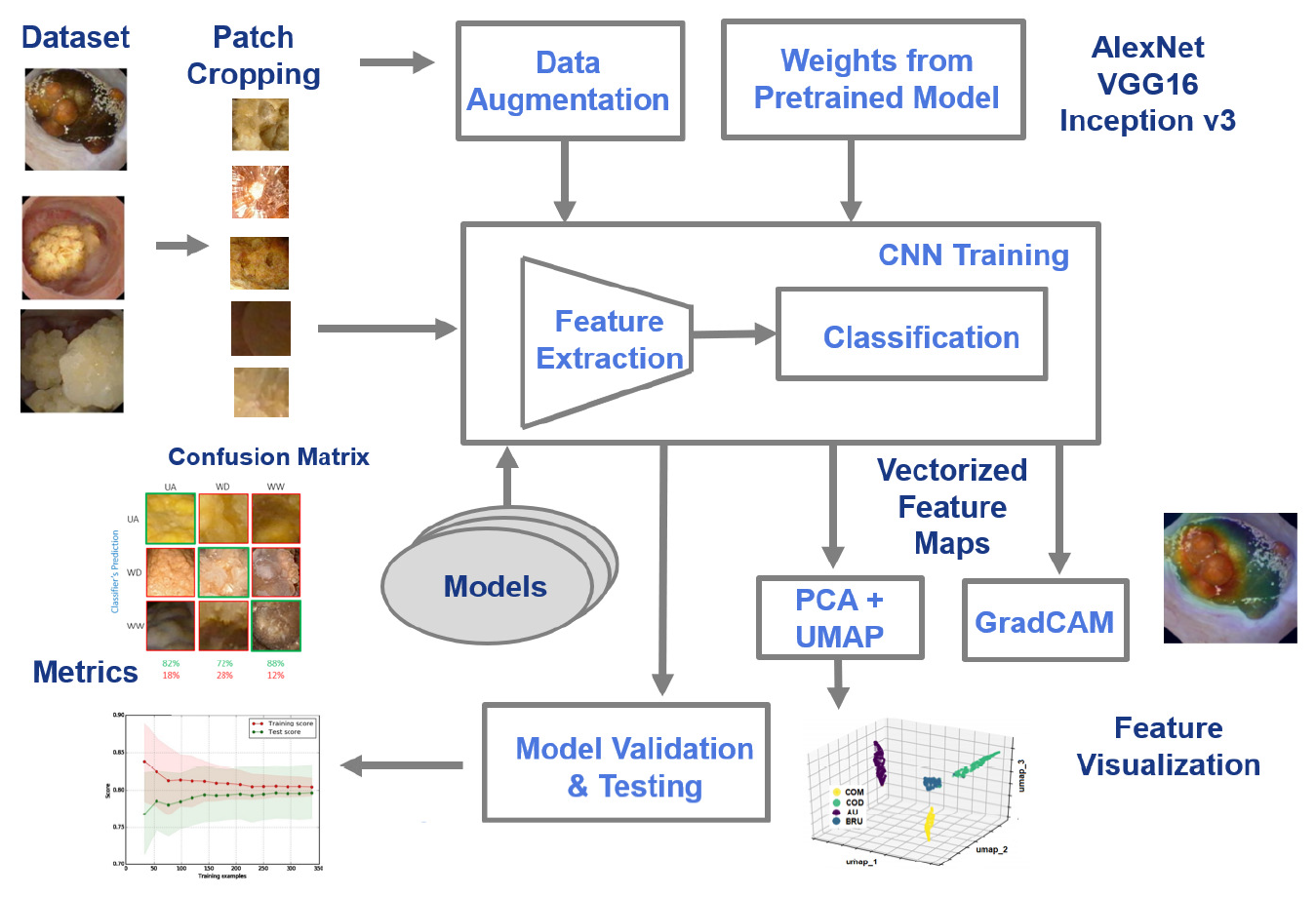}
\vspace{4pt}
\caption{Process followed for training the deep learning models: data augmentation and training using pre-trained feature extraction backbones, model selection and validation and feature visualization.}
\label{fig:diagram_deep}
\end{figure*}

The six chosen models represent a relatively large pool of shallow machine-learning methods. The  best hyper-parameter values for the combined surface and section patches descriptors are given below:  

\begin{itemize}
    \item{\textbf{SVM.} The best model was obtained by setting the C parameter value at 1.16, and by using a sigmoid kernel with its defaults values for the coeff0 (= 0.0) and gamma (= scale) hyper-parameters.}
    \item{\textbf{AdaBoost.} A set of decision tree classifiers was used for the model, the number of estimators and the maximum depth being equal to 100 and 12, respectively. The best LR was set to 0.1.}
    \item{\textbf{Bagging.} The bagging model has the same parameter values as AdaBoost, but the Random Forest three  was employed as the base estimator (the number of estimators equals 160).}
    \item{\textbf{Multi-Layer Perceptron (MLP).} The MLP model had the following hyper-parameter settings: it consists of three layers, and 200 neurons were used in the hidden later. Tt was trained for 200 epochs using the L-BFGS solver.}
    \item{\textbf{Random Forest (RF).} The RF  model with the best hyper-parameter settings  consists of 1800 estimators, a minimum split value of 5, a minimum of samples per leaf of 2, and a max depth of 50. Bootstrap was not used.}
    \item{\textbf{XGBoost.} The best hyperparemeter settings of the XGBoost model are the following.  The base score value was set to 0.5. {gbtree was used as booster. The learning rate was set to 0.1, while gamma value was 0. A maximum depth of 3 and 100 estimators were used.}}
\end{itemize}
\subsection{Deep-learning based classification methods.\label{deep_methods}}
 The efficiency  of deep-learning models lies on their ability to automatically extract highly discriminating features~\cite{litjens2017survey}. Section \textit{``Training of the chosen deep-learning models''} presents the deep learning models that were designed for the kidney stone classification. Then Section \textit{``Visualization of deep-feature data''} highlights the discrimination ability of the extracted deep-features.
\subsubsection{Training of the chosen deep-learning models. \label{dl_networks}}
Contrary to shallow machine learning solutions, the features extracted from images by deep-learning models do not correspond to a predefined physical information (e.g., relating to colors or textures), but depend on weight values linking the input data to class probabilities. After the model training, it is difficult to physically interpret the deep-features since they depend on numerous weights of the convolution layers. However, the appropriateness of deep-features to discriminate instances of different classes can be analysed after the learning phase using visualization tools like UMAP~\cite{mcinnes2018umap} or explainability techniques such as GradCAM~\cite{selvaraju2017grad}. The approach depicted in Fig.~\ref{fig:diagram_deep} was used to exploit various DL-models. Three DCNNs were first pre-trained using transfer learning, while the data (the patches whose generation was discussed in Section \textit{``Class balancing strategies''}) used for the final training were augmented.

CNN models with different extraction backbones and architectures (AlexNet, VGG16 and Inception v3) were tested in this contribution.  Due to the moderate amount of available data, the convolution layers of the DL-models were pre-trained using ImageNet. Moreover, to adapt the DL-models to the kidney stone recognition task, the fully connected (FC) layer of the feature extraction backbones was replaced by a custom FC layer consisting of 256 channels. The outputs of this layer are then concatenated with a  Batch Normalization module, followed by a ReLU activation function, another 256 channel FC layer and ends with a softmax layer with 4 class outputs for yielding the class prediction. The fully connected layers weights were randomly initialized. During the training of the three models, the weights in the convolutional layers (obtained during the pre-training with ImageNet) were maintained constant, and only the weights in the FC layers were updated.

Extensive data augmentation was performed in order to limit the overfitting induced by the small size of the training dataset. Additional patches were obtained by applying vertical and horizontal flips, perspective distortions, and four affine transformations on the patches extracted from the images. With this data augmentation,  the number of samples in the training set passed from 5,400 to 43,200 (10\%  of the samples  were hold out  for test  purposes). Further, the patch values were ``whitened’’ using~\eqref{eqWhitening}  in which the mean $m_i$ and standard deviation $\sigma_i$ of the colour values $P_i(x,y)$  are determined in each color channel:

\begin{equation} 
\label{eqWhitening}
    P_i^w(x,y)  =  \frac{P_i(x,y) – m_i}{\sigma_i}, \quad \textrm{with} \quad i = R, G, B.
\end{equation}

Each DL-model was  trained three times, i.e. with different datasets (only section patches, only surface patches and both patch types combined). Thus,  different parameter tuples were computed for each dataset which were classically split in three parts, namely training, validation and test sets. All the experimental studies reported in this paper made use of Pytorch 1.7.0 and CUDA 10.1. The hyper-parameters such as the learning rates  were automatically adjusted for each architecture using the optimizer provided by Pytorch (Lightning 1.0.2). LR values of 0.0001, 0.00005, and 0.0006 were obtained for AlexNet, VGG16, and Inception V3, respectively. The ADAM optimizer, a batch size of 64, and early stopping were employed in all tests.
\subsubsection{Visualization of deep-feature data.\label{deep_features}}
The result description in Section \textit{``Deep Learning using transfer learning''} is based on two visualization tools which enabled us to better understand the ability of DL-models to recognize kidney stone types using the  deep-features extracted from the three datasets (surface or/and section patches). 

%
%
\begin{table*}[t]
\centering
\begin{tabular}{|c|c|c|c|c|c|c|c|c|c|}
\hline
 & \multicolumn{3}{c|}{\textbf{Surface}} & \multicolumn{3}{c|}{{ \textbf{Section}}} & \multicolumn{3}{c|}{ \textbf{Mixed}} \\ \cline{2-10} 
\multirow{-2}{*}{{ \textbf{Classifier}}} & Precision &  Recall & F1 &  Precision &  Recall & F1 & 
Precision & Recall & F1 \\ \hline
\textbf{SVM} & 0.83 & 0.86 & 0.84 & 0.76 & \textit{0.86} & 0.80 & 0.79 & 0.77 & 0.78 \\ \hline
\textbf{AdaBoost} & 0.83 & 0.86 & 0.84 & 0.81 & 0.85 & \textit{0.83} & 0.81 & 0.81 & 0.81 \\ \hline
\textbf{Bagging} & 0.76 & 0.76 & 0.76 & 0.77 & 0.77 & 0.77 & 0.75 & 0.76 & 0.75 \\ \hline
\textbf{MLP} & 0.86 & \textit{0.91} & \textit{0.88} & 0.80 & 0.64 & 0.71 & 0.84 & 0.86 & 0.85 \\ \hline
\textbf{R. Forest} & \textit{0.87} & 0.82 & 0.84 & \textit{0.82} & 0.82 & 0.82 & \textit{0.91} & \textit{0.91} & \textit{0.91} \\ \hline
\textbf{XGBoost} & \textbf{0.93} & \textbf{0.93} & \textbf{0.93} & \textbf{0.89} & \textbf{0.89} & \textbf{0.89} & \textbf{0.96} & \textbf{0.96} & \textbf{0.96} \\ \hline
\end{tabular}
\caption{Comparison of the performance of six shallow machine learning models according to the data type (section or/and surface kidney stone patches). The numbers in bold and italics represent the best and the second best results, respectively.}
\label{ml_results}
\end{table*}
\begin{itemize}
    \item \textbf{UMAP} (Uniform Manifold Approximation and Projection for Dimension Reduction, \cite{mcinnes2018umap}) is used to construct a high dimensional (deep-learning) feature graph for each dataset type. The UMAP algorithm then reduces the dimensionality of the feature space by optimizing a low-dimensional graph so that both graphs are as structurally similar as possible. In this contribution, the deep-learning features are represented in a three dimensional space whose dimensions umap1, umap2 and umap3 give component values obtained after a non-linear dimension reduction. These 3D representations illustrate the class separability of the deep-features.
    \item \textbf{GradCAM} (gradient weighted class activation mapping, \cite{selvaraju2017grad}) Additionally, the features in the low-dimensional space are used for creating  heat maps. This visualization technique uses the class-specific gradient information flowing into the final convolutional layer of a CNN to produce a coarse localization map of the important patch regions which triggered the classifier output. GradCAM representations allow for a better understanding of some of the errors made both by shallow and deep learning-based models.  These class activation mappings can be used for determining the information (i.e.,  for finding the important color or texture features, or locating the image areas including the most important information) which favour a successful kidney stone recognition. 
\end{itemize}
\section{Results and discussion. \label{results}} 
Various experiments were carried out for evaluating both shallow machine learning-based and deep learning-based methods. In all, nine models were trained three times, namely i) solely with the section patches, ii) only with the surface patches, and iii) by mixing the two patches types. Each section or surface patch was classified by its dedicated model and by the model for mixed data.
\subsection{Experiments using shallow machine learning methods.\label{trad_methods}} 

This section compares the results of six shallow machine learning models which were tuned with the validation pipeline depicted in Fig.~\ref{fig:diagram_shallow}. All results are given for the complete 40 component  $HSV$/$LBP$ feature vectors when either only surface or only section information is used for the recognition, and  for the 80-component vector when both data types are simultaneously exploited for the classification.

When analysing Table~\ref{ml_results} it becomes clear that the first results given in the literature for ex-vivo data were significantly improved by  carefully choosing handcrafted features. Indeed, the authors in~\cite{serrat2017mystone} obtained a precision of 63\% (see Table~\ref{pworks2}) over four classes using a RF model exploiting $RGB$ color values and non optimized $LBP$ window sizes. As seen in Table~\ref{ml_results}, the precision (83\%, 76\%, and 91\% for surface patches, section patches and both patch types, respectively) is greatly improved when a RF model exploits $HSV$ color features and an appropriate $LBP$ window size. The results reported in Table~\ref{ml_results} for various shallow machine learning based methods outperform even the first results obtained with deep-learning approaches: in~\cite{torrell2018metric} the authors obtained a precision of 74\% over four classes using a Siamese CNN solution, while in~\cite{black2020deep} a ResNet 101 led to a precision from 71\% up to 94\% according to the class. 

It is also noticeable in Table~\ref{ml_results} that the performance criteria (precision, recall and F1-scores) exhibit different values for surface and section patches taken separately. On the one hand, surface patches are globally more discriminant than section patches, and two classifiers lead globally to the best (XGBoost) and second best (MLP) results for the precision, recall and F1-score values for surface data. On the other hand, XGBoost enables the best classification for section patches, whereas no second best classifier really emerges for this patch type (for the section patches, the second best criterion values were obtained by three different classifiers). This difference in an automated exploitation of surface and section data confirms the visual analysis of the expert who noticed in~\cite{estrade2022towards} that the classification of surface images is easier than that of section images, as the former present more texture information due to the crystallization process than the latter. However, it can be noticed that when using both surface and sections patches two classifier obtain systematically the best (XGBoost) and second best results (RF trees). These two shallow machine learning methods are able to reduce the effects of high intra-class variability and low inter-class differences when exploiting simultaneously surface and section data (the 80-component feature vector).The performances of the XGBoost and RF classifiers are compared in Section \textit{``Deep Learning using transfer learning''} with those of the tested deep-learning methods.
\subsection{Deep Learning using transfer learning. \label{results_deep}}
\begin{table*}[t!]
\centering
\begin{tabular}{|c|c|c|c|c|c|c|c|c|}
\hline
\multirow{2}{*}{\textbf{Method/ Class}} & \multicolumn{2}{c|}{\textbf{WW (Ia)}} & \multicolumn{2}{c|}{\textbf{WD (IIb)}} & \multicolumn{2}{c|}{\textbf{AU (IIIb)}} & \multicolumn{2}{c|}{\textbf{BRU (IVd)}} \\ \cline{2-9} 
 & Precision & Recall & Precision & Recall & Precision & Recall & Precision & Recall \\ \hline
\textbf{Random Forest} & 0.84 & 0.86 & 0.90 & \textit{0.95} & 0.88 & 0.67 & 0.90 & 0.92 \\ \hline
\textbf{XGBoost} & 0.92 & 0.96 & 0.91 & 0.91 & \textbf{0.97} & \textbf{0.96} & \textbf{0.96} & \textit{0.94} \\ \hline
\textbf{AlexNet} & 0.93 & \textbf{0.98} &\textbf{0.95} & 0.85 & 0.88 & \textit{0.92} & 0.93 & 0.92 \\ \hline
\textbf{VGG19} & \textit{0.97} & \textit{0.97} & 0.92 & 0.93 & 0.93 & 0.83 & 0.94 & 0.92 \\ \hline
\textbf{Inception v3} & \textbf{0.98} & \textit{0.97} & \textit{0.93} & \textbf{0.96} & \textit{0.95} & 0.90 & \textbf{0.96} & \textbf{0.95} \\ \hline
\end{tabular}
\caption{Precision and recall values obtained for the four classes with five classifiers. The best and second best criterion values are given for each class in bold and italics, respectively.}
\label{dl_results1}
\end{table*}
\begin{table*}[t!]
\centering
\begin{tabular}{|c|c|c|c|c|c|c|c|c|c|}
\hline
\multirow{2}{*}{\textbf{Method/ Class}} & \multicolumn{3}{c|}{\textbf{Surface}} & \multicolumn{3}{c|}{\textbf{Section}} & \multicolumn{3}{c|}{\textbf{Mixed}} \\ \cline{2-10} 
 & Precision & Recall & F1 & Precision & Recall & F1 & Precision & Recall & F1 \\ \hline
\textbf{Random Forest} & 0.87 & 0.82 & 0.84 & 0.82 & 0.82 & 0.82 & 0.91 & 0.91 & 0.91 \\ \hline
\textbf{XGBoost} & 0.92 & \textit{0.96} & 0.93 & 0.89 & 0.89 & 0.89 & \textit{0.96} & 0.96 & \textit{0.96} \\ \hline
\textbf{AlexNet} & 0.93 & 0.95 & 0.94 & 0.83 & 0.82 & 0.83 & 0.94 & \textit{0.97} & 0.95 \\ \hline
\textbf{VGG19} & \textit{0.95} & 0.94 & \textit{0.95} & \textit{0.91} & \textit{0.92} & \textit{0.92} & 0.95 & 0.96 & 0.95 \\ \hline
\textbf{Inception v3} & \textbf{0.98} & \textbf{0.97} & \textbf{0.97} & \textbf{0.94} & \textbf{0.96} & \textbf{0.95} & \textbf{0.97} & \textbf{0.98} & \textbf{0.97} \\ \hline
\end{tabular}
\caption{Weighted average precision, recall and F1-score comparison for the dataset including only surface patches, the dataset consisting only of section patches, and for the complete dataset with both patch types. The best and second best criterion values are given for each class in bold and italics, respectively}
\label{dl_results2}
\end{table*}

Tables~\ref{dl_results1} and~\ref{dl_results2} gather complementary results of both the two best shallow machine learning methods and the three tested deep-learning methods. Table~\ref{dl_results1} gives results for individual classes including the surface and section patches, whereas Table~\ref{dl_results2} provides class weighted performance criterion values for the surface patches, the section patches, and the patches of both types. 

It is noticeable that, among the deep-learning methods, only the Inception V3 model outperforms the best shallow machine learning method based on the XGBoost classifier. As seen in the last row of Table~\ref{dl_results1}, this DL-model exhibits the highest mean precision and recall values for all classes. This result is also confirmed in Table~\ref{dl_results2} for all patch datasets (section patches, surface patches and mixed patch types). It can also be observed in Table~\ref{dl_results2} that, similarly to the shallow machine learning methods, the simultaneous use of surface and section patches leads to the best results, whereas the three DL-based methods exhibit globally superior results when exploiting only surface images. This last observation confirms the visual classification results of the concordance study in~\cite{estrade2022towards}. However, among the DL-approaches, only Inception v3 has globally a higher precision (0.97) and recall (0.98) than the XGBoost classifier (0.96 for both the precision and the recall values).

The precision and recall values in bold (best result) and italics (second best result) given for each class in Table~\ref{dl_results1} were most often obtained by one of the three deep-learning models. Similarly, the deep-learning methods most often delivered the two best criteria values for the datasets including only one patch type (either surface patches or section patches in Table~\ref{dl_results2}). However, when considering globally the results (weighted criterion values for all patch types exploited together, see Table~\ref{dl_results2}), only the inception V3 model has slightly better results than the XGBoost model.

\begin{figure*} [b!] 
    \centering
    
    \subfloat[]{
    \includegraphics[width=0.35\linewidth]{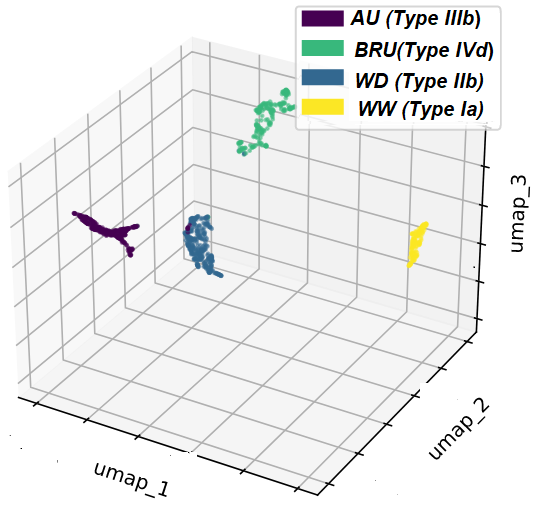}}
    \hspace{5mm}
    \subfloat[]{
    \includegraphics[width=0.35\linewidth]{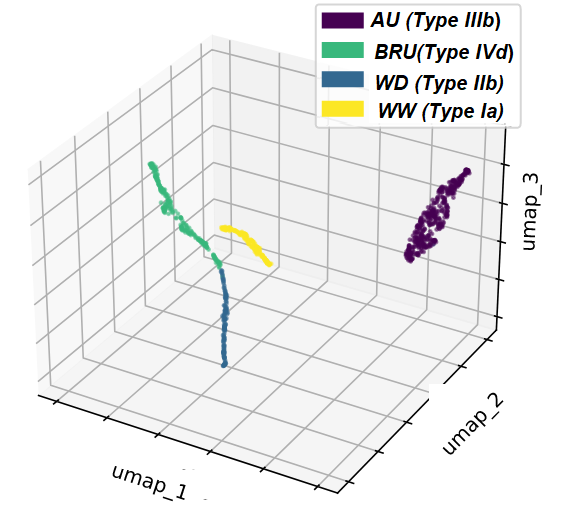}}
    
    \subfloat[]{
    \includegraphics[width=0.35\linewidth]{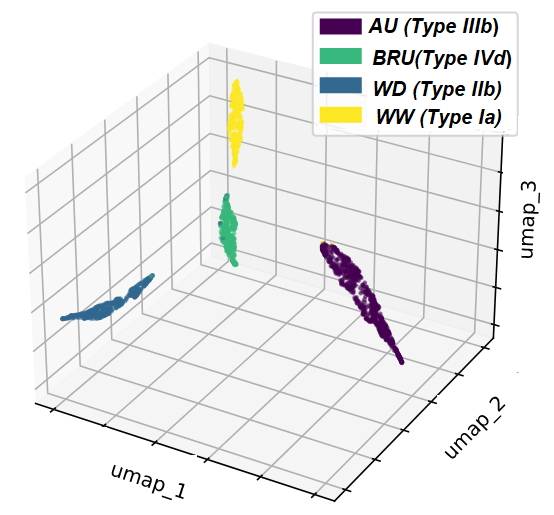}}
    \hspace{5mm}
    \subfloat[]{
    \includegraphics[width=0.35\linewidth]{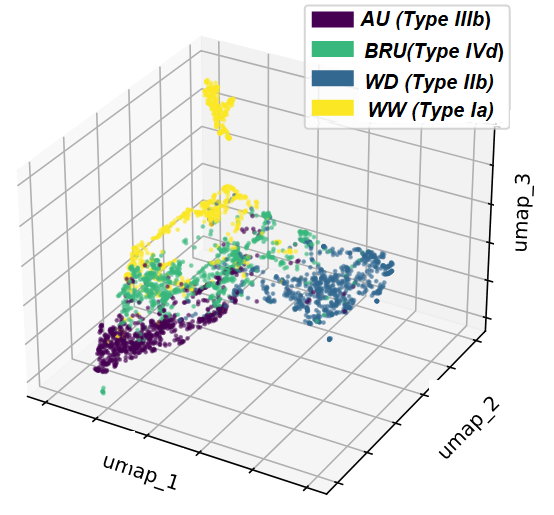}}
\vspace*{-3mm}
\caption{UMAP visualisation given for the AlexNet model. The values of the umap1, umap2, and umap3 components were obtained after a dimensionality reduction of the initial feature space. (a) Cluster representation of the four kidney stones classes when training the AlexNet model only with surface patches. The initial feature space (before dimensionality reduction) consisted only of deep-features extracted by the AlexNet convolutional layers. (b) Same as in (a), but by training the AlexNet model only with section patches. (c) UMAP visualisation for the feature extraction performed by training AlexNet with both section and surface patches. Although being elongated, the more distant clusters indicate improved classification performances in comparison to the clusters in (a) and (b). (d) The clusters obtained after a dimensionality reduction  of the  handcrafted $HSV/LBP$ feature space (same as in Fig. \ref{fig:umap}) are shown to allow for a comparison with the deep-feature clusters in (c). This comparison highlights the feature extraction improvement attained by a simple model such as AlexNet over traditional handcrafted features (here the three most discriminating components obtained  for the 80-component vectors for mixed section and surface patches).\label{fig:umap_deep}}
    \end{figure*}
The visualization of the three most discriminant UMAP components in Fig.~\ref{fig:umap_deep} explains why AlexNet exhibits rather moderate kidney stone recognition performances.
As it can be observed in  Fig.~\ref{fig:umap_deep}.(a), the feature extraction backbone of AlexNet determines deep-features that form compact and separated clusters for surface patches. This result is in agreement with the high recall values obtained for surface images in the concordance study in~\cite{estrade2022towards}. In contrast, the deep feature clusters produced by AlexNet for section patches are close and elongated. 
For instance, Fig.~\ref{fig:umap_deep}.(b) shows that the clusters corresponding to the weddellite and brushite classes are very close in the reduced UMAP feature space (the two clusters are even touching themselves in some reduced feature space places).The WD and WW clusters are also very close to each other, but without touching themselves. This explains why in Table~\ref{dl_results2} the precision and recall values are rather low  for section images, and high for surface images when using AlexNet.
 Fig.~\ref{fig:umap_deep}.(c) shows the feature clusters when training AlexNet on both surface and section patches. In this figure, the inter-cluster distances and compactness are visually similar to those of Fig.~\ref{fig:umap_deep}.(a). This observation is conform with the fact that the precision, recall and F1-score values of the surface dataset ($0.93\%$, $0.95\%$ and $0.94\%$ respectively, see Table~\ref{dl_results2}) are close to those of the mixed dataset ($0.94\%$, $0.97\%$ and $0.95\%$). While the section dataset taken separately leads to rather moderate recognition results, it improves slightly the classification results when it is associated to surface patches.
%


%
As seen in Table~\ref{dl_results2}, the observation made for Alexnet can be extended to the two other tested deep-learning methods: surface images are in general almost sufficient for obtaining a high classification performance, but mixing surface and section images increases the recall values of about $1\%$ in average (compare the recall values of AlexNet, VGG19 and Inception v3 with and without section patches).
For the sake of completeness, Fig.~\ref{fig:umap_deep} provides also a comparison between the dimensionality reduced deep-feature space (see Fig.~\ref{fig:umap_deep}.(c)) and the reduced handcrafted $HSV$-$LBP$ feature space determined with the surface and section datasets (see Fig.~\ref{fig:umap_deep}.(d)).  

When comparing these two 3D spaces, one could conclude that the class separability offered by deep-features leads to a more efficient classification than the handcrafted features, the inter-class distances and cluster compactness being visually the highest in Fig.~\ref{fig:umap_deep}.(c). However, XGBoost is able to exploit the less promising handcrafted features to achieve a classification with a performance slighly better than that of the VGG16 and Alexnet networks (see the F1-scores for mixed data in Table~\ref{dl_results2}).

\begin{figure}[t]
\centering
\includegraphics[width=\linewidth]{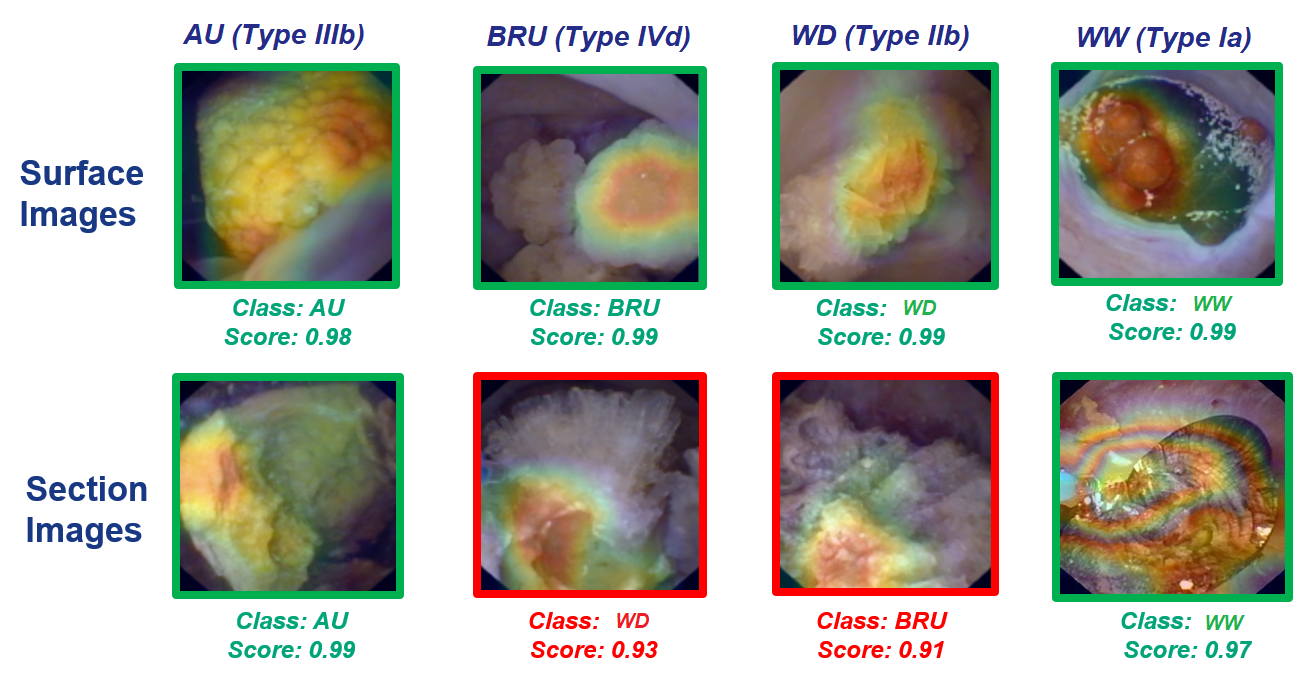}
\caption{Grad-CAM visualizations illustrating the performances of the AlexNet model. Complete kidney stone images are superimposed by their corresponding gradient-weighted class activation maps. The first and second rows correspond to the acid uric, brushite, whewellite and and wedelitte images acquired for the surface and section images, respectively. The green image frames indicate a correct classification by the GradCAM activation, while the red frames refer to as a very high activation score which led to  a classification error (the WW and BRU section images were misclassified).}
\label{fig:gradcam1}
\end{figure}

The performances of the AlexNet feature extraction back-bone are further discussed with the aid of the GradCAM network dissection and visualization method. In this visualization technique, the images are overlapped by  heatmaps in which color codes indicate the importance of particular image regions during the prediction making of deep learning models (from the red to the blue colors, the importance of the pixels decreases, while regions with grey-level values have no impact on the decision).  Fig.~\ref{fig:gradcam1} shows the activation maps for a surface (first row) and a section (second row) image of each class. The GradCAM visualisations were determined with models trained and tested on entire images of a unique patch type (either surface or section data).
From the surface images of first row, it can be inferred that the feature extraction backbone focuses in well-defined regions with significant presence of both color and texture information to produce highly discriminant features enabling a classification with high confidence (the green frames indicate a correct classification and the class score approaches 1). Trained clinicians also focus their attention on local and significant colour and texture images to recognized kidney stones either in microscopic images (during a morpho-constitutional analysis~\cite{daudon2012stone}) or in endoscopic data (during ureteroscopies~\cite{estrade2017pourquoi}).

However, it must to be noted  that GradCAM can produce relatively active heatmaps for images of a given class, even if the visualisation tool is fed with images from another class, as other components of the softmax layer vector might contribute to non-negligible class score values (for instance, the class score for the WD can be 0.75, whereas a value of 0.20 can be obtained for the brushite class and 0.025 for the other two classes). In such a situation, the precision, recall and F1-score values tend to be weak and the activation maps are sparser. For this reason, heatmaps should always be analysed by jointly considering the classification quality criteria.

\begin{figure*} [t!] 
    \centering
    \subfloat[]{
    \includegraphics[width=0.35\linewidth]{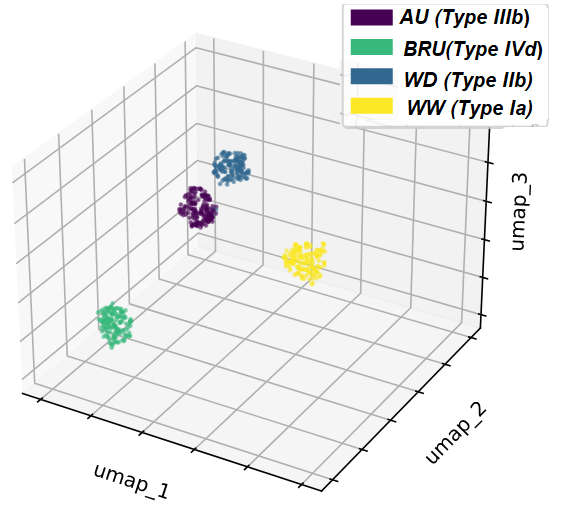}}
    \hspace{5mm}
    \subfloat[]{
    \includegraphics[width=0.35\linewidth]{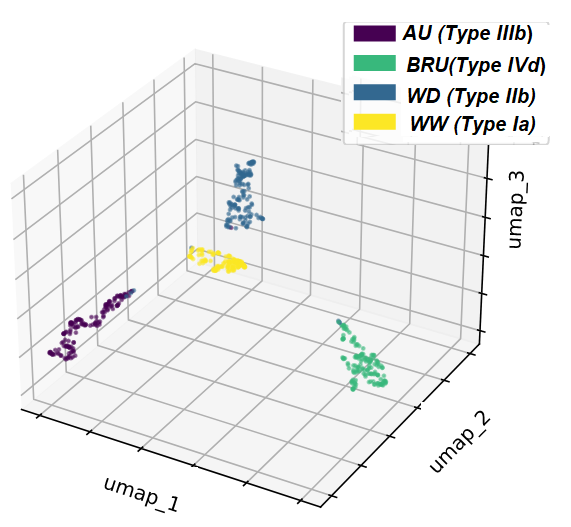}}
    \newline
    \subfloat[]{
    \includegraphics[width=0.35\linewidth]{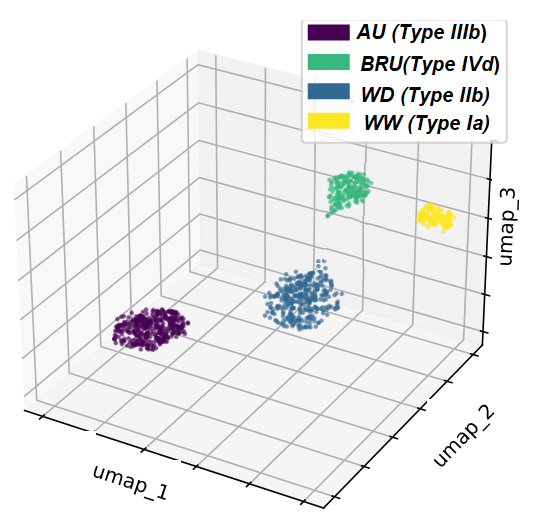}}
    \vspace*{-2mm}
    \caption{UMAP visualisation given for the Inception v3 model. (a) Representation of the four kidney stones classes when training the Inception v3 model only with surface patches. (b) Same cluster representation as in (a), but by training the model only with  section patches.  (c) Reduced feature space obtained by mixing both section and surface patches: the cluster separation is increased and leads to improved classification performances for the four classes.}
    \label{fig:umap_inception}
    \end{figure*}

This is shown in the second row of Fig.~\ref{fig:gradcam1} with GradCAMs of section images.  Even if the activation maps indicate a significant presence of colours and texture information, correct classification scores (close to 1) were only obtained for the acide urique and whewellite (COM) samples (0.99 and 0.97, respectively). In contrast, AlexNet misclassified the brushite and whedellite (WD) samples (assigning them the wrong label with a high confidence score of 0.93 and 0.91 scores, respectively). These results are in accordance with the UMAP visualizations in Fig.~\ref{fig:umap_deep} (b), where the clusters for COD and brushite classes are very close to each other. The Inception v3 deep-learning model exhibits the best overall performance (see Tables ~\ref{dl_results1} and ~\ref{dl_results2})  since it produce compact and distant clusters for all datasets  (surface, section or mixed patch types). The cluster separability of the Inception v3 backbone is illustrated in Fig.~\ref{fig:umap_inception} by the UMAP visualisation.

\section{Conclusion and perspectives.\label{conclusions}}

In this pilot study, it was shown that it is possible to train machine learning models (both shallow and deep learning-based) for recognizing the type of kidney stones using only digital images acquired with endoscopes during standard ureteroscopies.
The results presented in this contribution show that AI methods are potentially a precise solution to help urologists to recognize the morphology (i.e, the crystal type) of four kidney stone types (subgroups Ia, IIb, IIIb and IVd).  Additional works need to confirm this ability to identify the morphology on a larger number of pure and multilayered kidney stone classes. Until now, the FTIR analysis remains essential to complete the morphological analysis with the determination of  the biochemical composition of the stone. A solution to move towards a complete diagnosis during the endoscopy would be to equip ureteroscopy operating rooms with spectrometers which can collect IR signals in hollow organs using an optical fiber passing through the endoscope's operating channel. This contribution represents an important first step towards the immediate determination of an appropriate treatment avoiding recurrence in terms of kidney stone formation, while making the vaporization of kidney stones more systematic.

As thoroughly discussed, the kidney stones have various visual aspects that have been used  to propose taxonomies (based on color, texture and morphological descriptions) for aiding the urologist in their visual classification. Compared to the work by Serrat et al.~\cite{serrat2017mystone}, it was shown that a careful feature extraction and reduction can led to an efficient  kidney stone type recognition using shallow machine learning methods.  High performances were obtained when classifying the most common classes of urinary calculi (Ia, IIb, IIIb, IVd). For these four kidney stone types,  the XGBoost method led to an precision of 0.92 ,  0.89   and 0.96 when using only surface patches, solely section patches, and both patch types, respectively.  
Furthermore, this contribution is an extension of a previous preliminary study which only focused on shallow machine learning methods applied on a smaller class number. It was shown that some of the most common deep learning architectures (AlexNet, VGG16 and Inception v3) can be effectively trained for obtaining solutions with comparable or higher performances than  those obtained by Black et al.~\cite{black2020deep}. Some tested  CNN-models (e.g., Inception v3) are with a lower complexity than ResNet-101 and Restnet-152, but reach a slightly better precision due to their improved information density capabilities~\cite{canziani2016analysis}.  In this contribution, the weighted average precision obtained for Inception v3 equals  $[0.98, 0.94, 0.98]$ (for surface, section and mixed surface/section images, respectively), while high recall values were reached for all four used classes. However, the main difference between this study and previous works lies on the demonstration of the feasibility of classification methods making use of images acquired using flexible endosocopes under uncontrolled acquisition conditions (in previous studies such as~\cite{black2020deep}, the results were obtained in ex-vivo under ideal acquisition conditions).

In comparison to the most recent work that investigated the use of deep learning techniques for classifying  the morphology of different types of kidney stones also acquired in in-vivo~\cite{estrade2022towards}, the main contribution of this work lies in the thorough comparison of both shallow and deep learning architectures. This comparison also focused on the understanding of the features enabling a precise classification, as well as of the limitations of some methods. By comparing deep learning models of various levels of complexity, it was  graphically possible to confirm the results of the concordance study by Estrade et al.~\cite{estrade2022towards}. For instance, it was verified with various classification experiments and with the UMAP visualizations that both shallow and deep learning models can reach a high accuracy when classifying UA (Type IIIb kidney stones), but are less effective when images the of weddellite (Type IIb stones) and brushite (Type IVd) classes need to be distinguished. To a lesser extent, whewellite  images are also more complicate to be separated from the two previous classes. These observations explain the lower recall values for these three classes, both in the concordance study in~\cite{estrade2022towards} and in this contribution. Furthermore, the UMAP visualizations have been integrated into an interactive visualization tool~\cite{IvanGarcia} that enables the exploration of more complex models and databases (i.e, include more pure kidney stone classes or mixed stones for instance). It is also noticeable that the average precision obtained in~\cite{estrade2022towards} for surface and section images taken individually (0.94 in both cases, see Table~\ref{pworks2}) are lower than those obtained with an inception v3 architecture for surface images (0.98, see Table~\ref{dl_results2}) and mixed surface/section data (0.97).

The results presented in this paper show the potential and interest of AI methods to automate the determination of the causes (lithogenesis) of the kidney stone formation.  Nonetheless, additional tests should include other types of kidney stones with mixed composition to make an automated recognition procedure fully usable in clinical settings. Other  kidney stone types with a unique biochemical composition (as struvite and cystine) should also extend the database. Additionally, most works in the literature (including this one) make use of still images, which might limit the applicability of the computer vision systems proposed so far (the video sequences which are displayed on screens might be affected by motion blur and blood or debris can hide kidney stone parts). 

Other solutions that might be of interest to improve the classification results can be based on few shot learning approaches for object recognition and instance segmentation, the size of the available dataset  being relatively small. 
Also, when more stones types are included in the dataset, the proposed models might benefit from online or active learning techniques for adapting to new settings (for instance, kidney stones from people from countries with very different weather, an aspect that has not been studied so far). Furthermore, training deep learning models using images in other color spaces (as the $HSV$ or $HSI$ color spaces) is another promising area of research, as the obtained results can be more robust and smaller the deep-learning networks could be deployed, speeding up the inference time~\cite{anwer2021compact}. 

\section*{Acknowledgments}
The authors wish to acknowledge the Mexican Council for Science and Technology (CONACYT) for the support in terms of postgraduate scholarships in this project. We also thank the AI Hub and the Centro de Innovacion de Internet de las Cosas at Tecnologico de Monterrey for their support for carrying the experiments in this paper in their NVIDIA's DGX computer. 

\section*{Compliance with ethical approval}
The images were captured in medical procedures following the ethical principles outlined in the Helsinki Declaration of 1975, as revised in 2000, with the consent of the patients.

\bibliographystyle{cas-model2-names}

\bibliography{cas-refs}

\end{document}